\documentstyle[12pt]{article}

\begin{document}

\author{C.\ Bizdadea\thanks{%
e-mail address: bizdadea@hotmail.com}, I. Negru and S.\ O.\ Saliu\thanks{%
e-mail addresses: osaliu@central.ucv.ro or odile\_saliu@hotmail.com} \\
Department of Physics, University of Craiova\\
13 A.\ I.\ Cuza Str., Craiova R-1100, Romania}
\title{Irreducible antifield BRST-anti-BRST formalism for reducible
gauge theories}
\maketitle

\begin{abstract}
In this paper we develop an irreducible antifield
BRST-anti-BRST formalism for reducible gauge
theories. 

\end{abstract}

\section{Introduction}
\noindent
The most powerful manifestly covariant quantization method for gauge
theories was proved to be the antifield BRST formalism \cite{1}--\cite{5}.
An other approach of the same kind is represented by the antifield
BRST-anti-BRST formalism. The BRST-anti-BRST method was differently
implemented at the Hamiltonian \cite{6}--\cite{10} and Lagrangian \cite{10}--%
\cite{23} levels. Although it does not play such an important role like the
BRST symmetry itself, the BRST-anti-BRST formulation is a helpful background
for the geometrical (superfield) description of the BRST transformation, the
investigation of the perturbative renormalizability of the Yang-Mills
models, a consistent approach of anomalies, as well as for the correct
understanding of the non-minimal sector involved with the BRST quantization 
\cite{24}--\cite{30}. The antifield BRST-anti-BRST symmetry can be
implemented in the context of irreducible, as well as of reducible gauge
theories. However, in the reducible case the BRST-anti-BRST ghost and
antifield spectra are more involved precisely due to the fact that the gauge
generators are no longer independent. In view of this, the derivation of the
solution to the master equation corresponding to the reducible case is more
difficult than in the irreducible one.

This paper investigates the possibility of quantizing reducible gauge
theories by employing the BRST-anti-BRST prescriptions for irreducible
systems. Our main result consists in proving that a large class of reducible
gauge systems can be covariantly quantized accordingly the irreducible
antifield BRST-anti-BRST manner. Our treatment is based on the fact that the
antifield BRST-anti-BRST symmetry for a given gauge theory exists simply
provided that the corresponding antifield BRST symmetry exists. In this
light, we enforce the following steps: (i) we transform the initial
reducible gauge theory into an irreducible one in a manner that allows the
replacement of the BRST quantization of the reducible system by that of the
irreducible theory, and (ii) we quantize the irreducible gauge theory within
the antifield BRST-anti-BRST framework. Step (i) results in the possibility
to derive an irreducible BRST symmetry associated with the reducible theory.
This makes legitimate the application of the irreducible antifield
BRST-anti-BRST machinery to the irreducible theory associated with the
original reducible system.

Our paper is organized in five sections. Section 2 realizes a brief review
on the main ingredients of the irreducible antifield BRST-anti-BRST
construction. In Section 3 we derive an irreducible theory associated with
the starting reducible one and show that it is permissible from the BRST
point of view to replace the quantization of the reducible system by that of
the irreducible theory. We subsequently develop the antifield BRST-anti-BRST
quantization of the irreducible theory associated with the reducible system.
Section 4 illustrates the theoretical part of the paper for the
Freedman-Townsend model and for a model with abelian three-form gauge fields.
Section 5 ends the paper with some conclusions.

\section{Main ideas of the irreducible antifield BRST-anti-BRST construction}
\noindent
In this section we present a summary of basic elements required at the
construction of the BRST-anti-BRST symmetry corresponding to an irreducible
gauge theory. We start from an arbitrary action (local functional) depending
on the fields $\Phi ^i$%
\begin{equation}
\label{2.1}S_0^L\left[ \Phi ^i\right] =\int d^DxL\left( \Phi ^i(x),\partial
_{\mu _1}\Phi ^i(x),\cdots ,\partial _{\mu _1}\cdots \partial _{\mu _s}\Phi
^i(x)\right) , 
\end{equation}
which is assumed invariant under the gauge transformations (written in the
De Witt condensed notations) 
\begin{equation}
\label{2.2}\delta _\epsilon \Phi ^i=R_{\;\;\alpha }^i\epsilon ^\alpha \quad
\left( \Leftrightarrow \delta _\epsilon \Phi ^i(x)=\int d^DyR_{\;\;\alpha
}^i(x,y)\epsilon ^\alpha (y)\right) . 
\end{equation}
For definiteness we consider the bosonic case, but the analysis can be
straightforwardly extended to fermions modulo introducing some appropriate
phases. We suppose that $R_{\;\;\alpha }^i$ form an irreducible generating
set, i.e., these functions are independent and complete. The completeness of
the gauge generators induces that 
\begin{equation}
\label{2.3}R_{\;\;\alpha }^j\frac{\delta R_{\;\;\beta }^i}{\delta \Phi ^j}%
-R_{\;\;\beta }^j\frac{\delta R_{\;\;\alpha }^i}{\delta \Phi ^j}\approx
C_{\;\;\alpha \beta }^\gamma R_{\;\;\gamma }^i, 
\end{equation}
where $C_{\;\;\alpha \beta }^\gamma $ may involve the fields, and the weak
equality `$\approx $' means an equality valid when the field equations hold.

The fundamental scope of the antifield BRST-anti-BRST formalism is to
construct two differentials defining an algebra of the type 
\begin{equation}
\label{2.4}s_1^2=0=s_2^2,\;s_1s_2+s_2s_1=0, 
\end{equation}
where $s_1$ and $s_2$ are respectively called the BRST and anti-BRST
operators, and must be such their cohomology at degree zero is given by the
classical observables of (\ref{2.1}) (gauge invariant functions defined on
the stationary surface of field equations $\Sigma :\frac{\delta S_0^L}{%
\delta \Phi ^i}=0$). From (\ref{2.4}) and the zeroth order cohomological
requirement it is clear that 
\begin{equation}
\label{2.5}s=s_1+s_2, 
\end{equation}
is also nilpotent, and, moreover, describes a BRST symmetry associated with
a redundant description of the gauge symmetries inferred by duplicating the
gauge generators. Conversely, any such redundant BRST symmetry of (\ref{2.1}%
) that splits like in (\ref{2.5}) implies the BRST-anti-BRST algebra (\ref
{2.4}) for its separate pieces plus the isomorphism between the zero degree
cohomologies of the individual parts and the classical observables of (\ref
{2.1}). These two aspects lead to the fact that one can replace (\ref{2.4})
by the unique relation 
\begin{equation}
\label{2.6}s^2=0, 
\end{equation}
provided $s$ can be made to split as in (\ref{2.5}). In other words, one can
simply follow the standard BRST rules for first-stage reducible gauge
systems in order to construct $s$ (and therefore $s_1$ and $s_2$, once the
split is shown to hold). In order to handle appropriately the two pieces of $%
s$ it is necessary to introduce a bidegree that distangles between them. It
is called ghost bidegree or bighost number, is denoted by $bigh=(gh_1,gh_2)$%
, and is defined through 
\begin{equation}
\label{2.7}bigh(s_1)=(1,0)\;,bigh(s_2)=(0,1), 
\end{equation}
such that the resulting degree, called ghost number and denoted by $gh$,
will be $gh=gh_1+gh_2$.

In order to perform a proper construction of $s$ (and consequently of $s_1$
and $s_2$), it is necessary to duplicate the gauge generators $R_{\;\;\alpha
}^i$ and to introduce the corresponding reducibility functions 
\begin{equation}
\label{2.8}\left( 
\begin{array}{c}
\delta _{\;\;\beta }^\alpha \\
-\delta _{\;\;\beta }^\alpha
\end{array}
\right) . 
\end{equation}
Following the standard BRST receipt, the ghost spectrum will contain the
ghosts $\eta _1^\alpha $, $\eta _2^\alpha $ and the ghosts of ghosts $\pi
^\alpha $, respectively associated with the new gauge generators $%
(R_{\;\;\alpha }^i,R_{\;\;\alpha }^i)$ and the reducibility functions (\ref
{2.8}).The bighost number and Grassmann parity ($\epsilon $) of the original
fields and ghost spectrum variables are defined by setting 
\begin{equation}
\label{2.9}bigh(\Phi ^i)=(0,0),bigh(\eta _1^\alpha )=(1,0),bigh(\eta
_2^\alpha )=(0,1),bigh(\pi ^\alpha )=(1,1), 
\end{equation}
\begin{equation}
\label{2.10}\epsilon (\Phi ^i)=0,\;\epsilon (\eta _1^\alpha )=\epsilon (\eta
_2^\alpha )=1,\;\epsilon (\pi ^\alpha )=0, 
\end{equation}
such that $gh(\eta _1^\alpha )=gh(\eta _2^\alpha )=1$ and $gh(\pi ^\alpha
)=2 $. From now on we include the ghost bidegree of an object by means of an
additional superscript such that if $F$ has $bigh(F)=(a,b)$, we simply write 
$\stackrel{(a,b)}{F}$. For further purpose, we generically call the original
fields, ghosts and ghosts of ghosts by `fields', and denote them like 
\begin{equation}
\label{2.11}\Phi ^A=\left( \stackrel{(0,0)}{\Phi ^i},\stackrel{(1,0)}{\eta
_1^\alpha },\stackrel{(0,1)}{\eta _2^\alpha },\stackrel{(1,1)}{\pi ^\alpha }%
\right) . 
\end{equation}
The longitudinal exterior derivative associated with the redundant
description of the gauge symmetries, $D$, is defined on the stationary
surface $\Sigma $ accordingly the usual BRST prescriptions as 
\begin{equation}
\label{2.12}D\stackrel{(0,0)}{\Phi ^i}=R_{\;\;\alpha }^i\left( \stackrel{%
(1,0)}{\eta _1^\alpha }+\stackrel{(0,1)}{\eta _2^\alpha }\right) , 
\end{equation}
\begin{equation}
\label{2.13}D\stackrel{(1,0)}{\eta _1^\alpha }=\stackrel{(1,1)}{\pi ^\alpha }%
+\frac 12C_{\;\;\beta \gamma }^\alpha \stackrel{(1,0)}{\eta _1^\beta }\left( 
\stackrel{(1,0)}{\eta _1^\gamma }+\stackrel{(0,1)}{\eta _2^\gamma }\right) , 
\end{equation}
\begin{equation}
\label{2.14}D\stackrel{(0,1)}{\eta _2^\alpha }=-\stackrel{(1,1)}{\pi ^\alpha 
}+\frac 12C_{\;\;\beta \gamma }^\alpha \stackrel{(0,1)}{\eta _2^\beta }%
\left( \stackrel{(1,0)}{\eta _1^\gamma }+\stackrel{(0,1)}{\eta _2^\gamma }%
\right) , 
\end{equation}
\begin{equation}
\label{2.15}D\stackrel{(1,1)}{\pi ^\alpha }=\frac 12C_{\;\;\beta \gamma
}^\alpha \stackrel{(1,1)}{\pi ^\beta }\left( \stackrel{(1,0)}{\eta _1^\gamma 
}+\stackrel{(0,1)}{\eta _2^\gamma }\right) , 
\end{equation}
such that $D^2\approx 0$. Thus, the total differential $D$ splits as $%
D=D_1+D_2$, with $bigh(D_1)=(1,0)$ and $bigh(D_2)=(0,1)$. The action of $D_1$
and $D_2$ can be immediately inferred from (\ref{2.12}--\ref{2.15}) by
identifying the components accordingly the ghost bidegree. The weak
nilpotency of $D$ yields $D_1^2\approx 0\approx D_2^2$ and $%
D_1D_2+D_2D_1\approx 0$.

It is well-known that in the standard BRST formalism there is one
antibracket with ghost number one. Here it is necessary to construct two
antibrackets in order to make the antibracket structure compatible with the
ghost bidegrees of the ghost spectrum while preserving the symmetry between
the two degrees $gh_1$ and $gh_2$. We denote the two antibrackets by $\left(
,\right) _1$ and $\left( ,\right) _2$ requiring that they possess the
bighost number $(1,0)$, respectively, $(0,1)$, and introduce a pair of
antifields $(\Phi _A^{*(1)},\Phi _A^{*(2)})$ respectively conjugated to a
field $\Phi ^A$ in the first and second antibracket. The characteristics of
the antifields are as follows 
\begin{equation}
\label{2.20}bigh\left( \Phi _A^{*(1)}\right) =\left( -gh_1(\Phi
^A)-1,-gh_2(\Phi ^A)\right) , 
\end{equation}
\begin{equation}
\label{2.21}bigh\left( \Phi _A^{*(2)}\right) =\left( -gh_1(\Phi
^A),-gh_2(\Phi ^A)-1\right) , 
\end{equation}
\begin{equation}
\label{2.22}\epsilon \left( \Phi _A^{*(1)}\right) =\epsilon \left( \Phi
_A^{*(2)}\right) =\epsilon (\Phi ^A)+1\;{\rm mod}\;2, 
\end{equation}
and the fundamental antibrackets are defined by 
\begin{equation}
\label{2.23}\left( \Phi ^A,\Phi _B^{*(1)}\right) _1=\left( \Phi ^A,\Phi
_B^{*(2)}\right) _2=\delta _{\;\;B}^A,\left( \Phi ^A,\Phi _B^{*(1)}\right)
_2=\left( \Phi ^A,\Phi _B^{*(2)}\right) _1=0. 
\end{equation}
However, there appear two problems with the definition of the Koszul-Tate
operator along the standard line of the antifield procedure, namely, it
fails to be nilpotent and there are non trivial co-cycles at positive
resolution degrees. These matters are due in principle to the fact that at
the Lagrangian level the duplication of the gauge symmetries is not
accompanied by a duplication of the field equations defining the stationary
surface. Therefore, the complete reducible description of gauge orbits does
not define a complete redundant description of the stationary surface. These
points prevent the direct construction of a Koszul-Tate differential $\delta 
$ that splits into two pieces generating a biresolution of $C^\infty (\Sigma
)$ in the usual manner. The difficulties mentioned above can be surpassed by
adding further variables $\bar \Phi _A$ called `bar variables', with the
features 
\begin{equation}
\label{2.24}bigh\left( \bar \Phi _A\right) =\left( -gh_1\Phi ^A-1,-gh_2\Phi
^A-1\right) ,\;\epsilon \left( \bar \Phi _A\right) =\epsilon (\Phi ^A), 
\end{equation}
and by modifying the action of $\delta $ as the sum between a canonical
(BRST-like component) $\delta _{{\rm can}}$ and a non-canonical part
expressed via an operator $V$ that acts (only on the bar variables) through 
\begin{equation}
\label{2.25}V=\left( \Phi _A^{*(2)}-\Phi _A^{*(1)}\right) \frac{\vec \delta 
}{\delta \bar \Phi _A}, 
\end{equation}
such that 
\begin{equation}
\label{2.25a}\delta =\delta _{{\rm can}}+V, 
\end{equation}
with $gh(\delta )=gh(\delta _{{\rm can}})=gh(V)=1$. Putting together the
generators $\Phi _A^{*(1)}$, $\Phi _A^{*(2)}$ and $\bar \Phi _A$ of $\delta $%
, we find the BRST-anti-BRST antifield spectrum for an original irreducible
gauge theory under the form 
\begin{equation}
\label{2.26}\Phi _A^{*(1)}=\left( \stackrel{(-1,0)}{\Phi }_i^{*(1)},%
\stackrel{(-2,0)}{\eta }_\alpha ^{*(11)},\stackrel{(-1,-1)}{\eta }_\alpha
^{*(12)},\stackrel{(-2,-1)}{\pi }_\alpha ^{*(1)}\right) , 
\end{equation}
\begin{equation}
\label{2.27}\Phi _A^{*(2)}=\left( \stackrel{(0,-1)}{\Phi }_i^{*(2)},%
\stackrel{(-1,-1)}{\eta }_\alpha ^{*(21)},\stackrel{(0,-2)}{\eta }_\alpha
^{*(22)},\stackrel{(-1,-2)}{\pi }_\alpha ^{*(2)}\right) , 
\end{equation}
\begin{equation}
\label{2.28}\bar \Phi _A=\left( \stackrel{(-1,-1)}{\bar \Phi }_i,\stackrel{%
(-2,-1)}{\bar \eta }_{1\alpha },\stackrel{(-1,-2)}{\bar \eta }_{2\alpha },%
\stackrel{(-2,-2)}{\bar \pi }_\alpha \right) . 
\end{equation}
The correct actions of the Koszul-Tate operator $\delta $ associated with
the redundant description of the gauge symmetries read as 
\begin{equation}
\label{2.28a}\delta \Phi ^A=0, 
\end{equation}
\begin{equation}
\label{2.29}\delta \stackrel{(-1,0)}{\Phi }_i^{*(1)}=\delta \stackrel{(0,-1)%
}{\Phi }_i^{*(2)}=-\frac{\delta S_0^L}{\delta \Phi ^i}, 
\end{equation}
\begin{equation}
\label{2.30}\delta \stackrel{(-2,0)}{\eta }_\alpha ^{*(11)}=\delta \stackrel{%
(-1,-1)}{\eta }_\alpha ^{*(21)}=\stackrel{(-1,0)}{\Phi }_i^{*(1)}R_{\;\;%
\alpha }^i, 
\end{equation}
\begin{equation}
\label{2.31}\delta \stackrel{(-1,-1)}{\eta }_\alpha ^{*(12)}=\delta 
\stackrel{(0,-2)}{\eta }_\alpha ^{*(22)}=\stackrel{(0,-1)}{\Phi }%
_i^{*(2)}R_{\;\;\alpha }^i, 
\end{equation}
\begin{equation}
\label{2.32}\delta \stackrel{(-2,-1)}{\pi }_\alpha ^{*(1)}=\delta \stackrel{%
(-1,-2)}{\pi }_\alpha ^{*(2)}=-\left( \stackrel{(-1,-1)}{\eta }_\alpha
^{*(21)}-\stackrel{(-1,-1)}{\eta }_\alpha ^{*(12)}+\stackrel{(-1,-1)}{\bar
\Phi }_iR_{\;\;\alpha }^i\right) , 
\end{equation}
\begin{equation}
\label{2.33}\delta \bar \Phi _A=\Phi _A^{*(2)}-\Phi _A^{*(1)}. 
\end{equation}
Under these considerations, the total Koszul-Tate differential splits as $%
\delta =\delta _1+\delta _2$, with $bigh(\delta _1)=(1,0)$, $bigh(\delta
_2)=(0,1)$, the concrete actions of the two Koszul-Tate components being
deduced on account of identifying the expressions from (\ref{2.28a}--\ref
{2.33}) with respect to the resolution bidegree (also called antighost
bidegree or biantighost number). The prior definitions restore both the
nilpotency and the acyclicity of $\delta $ at positive resolution degrees,
and, moreover, induce the required relations 
\begin{equation}
\label{2.40}\delta _1^2=0=\delta _2^2,\delta _1\delta _2+\delta _2\delta
_1=0. 
\end{equation}
In addition, it is easy to check that while $\delta $ realizes a resolution
of the algebra $C^\infty (\Sigma )$, $\left( \delta _1,\delta _2\right) $
perform a biresolution of the same algebra. The resolution bidegree $%
bires=(res_1,res_2)$ of the BRST and anti-BRST Koszul-Tate components are
expressed as expected by $bires(\delta _1)=(-1,0)$, $bires(\delta _2)=(0,-1)$%
, while the resolution degree of an object with the resolution bidegree $%
(res_1,res_2)$ is obtained via $res=res_1+res_2$. The resolution bidegrees
of all the `fields' are $(0,0)$, and those of the antifields (\ref{2.26}--%
\ref{2.28}) can be computed by $bires\left( \stackrel{\left(
gh_1,gh_2\right) }{\rm antifield}\right) =\left( -gh_1,-gh_2\right) $. Now,
it is easy to see that $\delta _1$ and $\delta _2$ can be decomposed like $%
\delta $ in (\ref{2.25a}), i.e., 
\begin{equation}
\label{2.41}\delta _1=\delta _{1{\rm can}}+V_1,\;\delta _2=\delta _{2{\rm can%
}}+V_2, 
\end{equation}
with $bires(\delta _{1{\rm can}})=(-1,0)=bires(V_1)$, $bires(\delta _{2{\rm %
can}})=(0,-1)=bires(V_2)$, where 
\begin{equation}
\label{2.42}V_1=\Phi _A^{*(2)}\frac{\vec \delta }{\delta \bar \Phi _A}%
,\;V_2=-\Phi _A^{*(1)}\frac{\vec \delta }{\delta \bar \Phi _A}. 
\end{equation}

In the standard antifield approach the implementation of the BRST symmetry
is accomplished by means of a ghost number zero bosonic anticanonical
generator that is the solution of the master equation and leads to the
effective action. In the antifield BRST-anti-BRST approach there is also a
single bosonic generator $S$ of ghost bidegree $(0,0)$ (and thus of ghost
number zero) that implements this symmetry through 
\begin{equation}
\label{2.49}sF=\left( F,S\right) +VF, 
\end{equation}
for any $F$ depending on the `fields' and antifields, where the antibracket $%
\left( ,\right) $ is defined by 
\begin{equation}
\label{2.49a}\left( ,\right) =\left( ,\right) _1+\left( ,\right) _2. 
\end{equation}
The nilpotency of $s$ implies at the antibracket level the classical master
equation of the BRST-anti-BRST formalism 
\begin{equation}
\label{2.50}\frac 12\left( S,S\right) +VS=0, 
\end{equation}
whose generator is subject to the boundary conditions 
\begin{equation}
\label{2.46}\stackrel{(0)}{S}=S_0^L,\;\stackrel{(1)}{S}=\stackrel{(-1,0)}{%
\Phi }_i^{*(1)}R_{\;\;\alpha }^i\stackrel{(1,0)}{\eta _1^\alpha }+\stackrel{%
(0,-1)}{\Phi }_i^{*(2)}R_{\;\;\alpha }^i\stackrel{(0,1)}{\eta _2^\alpha }, 
\end{equation}
\begin{equation}
\label{2.47}\stackrel{(2)}{S}=\left( \stackrel{(-1,-1)}{\eta }_\alpha
^{*(21)}-\stackrel{(-1,-1)}{\eta }_\alpha ^{*(12)}+\stackrel{(-1,-1)}{\bar
\Phi }_iR_{\;\;\alpha }^i\right) \stackrel{(1,1)}{\pi ^\alpha }+{\rm ``more"}%
, 
\end{equation}
and to the properties $\epsilon (S)=0$, $bigh(S)=(0,0)$. Actually, it was
shown \cite{10} that the solution to the master equation (\ref{2.50})
satisfying the required conditions exists, which further yields that we can
decompose $s$ precisely into two pieces $s_1$ and $s_2$ of ghosts bidegrees (%
\ref{2.7}) that act like 
\begin{equation}
\label{2.51}s_aF=\left( F,S\right) _a+V_aF,\;a=1,2, 
\end{equation}
where $\left( ,\right) _a$ and $V_a$ are defined by (\ref{2.23}) and (\ref
{2.42}). The master equation (\ref{2.50}) is in fact equivalent with two
equations, namely, 
\begin{equation}
\label{2.52}\frac 12\left( S,S\right) _1+V_1S=0,\;\frac 12\left( S,S\right)
_2+V_2S=0. 
\end{equation}
However, the results obtained are not entirely convenient as the generator
of the Lagrangian BRST-anti-BRST symmetry is neither BRST, nor anti-BRST
invariant. This matter will be solved within the gauge-fixing process.

In order to appropriately fix the gauge, it is necessary to forget for the
moment about the second antibracket, and to add some new fields in the
theory, denoted by $\mu _{(1)}^A$ and $\rho _{(1)}^A$, which are
respectively conjugated in the first antibracket with $\overline{\Phi }_A$
and $\Phi _A^{*(2)}$ 
\begin{equation}
\label{2.53}\left( \mu _{(1)}^A,\bar \Phi _B\right) _1=\delta _B^A, 
\end{equation}
\begin{equation}
\label{2.54}\left( \Phi _A^{*(2)},\rho _{(1)}^B\right) _1=\delta _A^B. 
\end{equation}
The properties of the new variables read as 
\begin{equation}
\label{2.55}bigh\left( \mu _{(1)}^A\right) =\left( -gh_1\bar \Phi
_A-1,-gh_2\bar \Phi _A\right) ,\;\epsilon \left( \mu _{(1)}^A\right)
=\epsilon \left( \bar \Phi _A\right) +1\;{\rm mod}\;2, 
\end{equation}
\begin{equation}
\label{2.56}bigh\left( \rho _{(1)}^A\right) =\left( -gh_1\Phi
_A^{*(2)}-1,-gh_2\Phi _A^{*(2)}\right) ,\;\epsilon \left( \rho
_{(1)}^A\right) =\epsilon \left( \Phi _A^{*(2)}\right) +1\;{\rm mod}\;2. 
\end{equation}
The generator of the BRST-anti-BRST symmetry is extended on the new
variables by means of the relation 
\begin{equation}
\label{2.57}S_1=S+\Phi _A^{*(2)}\mu _{(1)}^A, 
\end{equation}
such that the master equation (\ref{2.50}) will be equivalent to $\left(
S_1,S_1\right) _1=0$, which has the well-known form of the master equation
from the standard antifield BRST method. Applying now the usual BRST
gauge-fixing process, namely, choosing a gauge-fixing fermion that involves
only the variables playing role of fields $\psi =\psi \left[ \Phi ^A,\mu
_{(1)}^A,\Phi _A^{*(2)}\right] $, we can eliminate the variables playing
role of antifields from the theory and find as usually the corresponding
gauge-fixed action, $S_{1_\psi }$. An alternative way of fixing the gauge is
to take a bosonic functional $F$ that depends only on the `fields' $\Phi ^A$
through 
\begin{equation}
\label{2.61}\psi =\mu _{(1)}^A\frac{\stackrel{\leftarrow }{\delta }F}{\delta
\Phi ^A}, 
\end{equation}
which yields 
\begin{equation}
\label{2.62}\bar \Phi _A=\frac{\stackrel{\leftarrow }{\delta }F}{\delta \Phi
^A},\;\Phi _B^{*(1)}=\mu _{(1)}^A\frac{\stackrel{\leftarrow }{\delta }^2F}{%
\delta \Phi ^B\delta \Phi ^A},\;\rho _{(1)}^A=0. 
\end{equation}
Introducing some Lagrange multipliers $\mu _{(2)}^A$ and $\lambda ^A$ that
implement the gauge conditions (\ref{2.62}), we obtain that the usual path
integral 
\begin{equation}
\label{2.63}Z_\psi =\int {\cal D}\Phi ^A{\cal D}\mu _{(1)}^A{\cal D}\Phi
_A^{*(2)}\exp iS_{1_\psi }, 
\end{equation}
becomes 
\begin{equation}
\label{2.64}Z_F=\int {\cal D}\Phi ^A{\cal D}\mu _{(1)}^A{\cal D}\Phi
_A^{*(1)}{\cal D}\mu _{(2)}^A{\cal D}\Phi _A^{*(2)}{\cal D}\lambda ^A{\cal D}%
\bar \Phi _A\exp iS_F, 
\end{equation}
with the effective action $S_F$ given by 
\begin{equation}
\label{2.65}S_F=S+\Phi _A^{*(2)}\mu _{(1)}^A-\Phi _A^{*(1)}\mu _{(2)}^A+\mu
_{(1)}^A\frac{\stackrel{\leftarrow }{\delta }^2F}{\delta \Phi ^B\delta \Phi
^A}\mu _{(2)}^B+\left( \bar \Phi _A-\frac{\stackrel{\leftarrow }{\delta }F}{%
\delta \Phi ^A}\right) \lambda ^A. 
\end{equation}
We remark that if one integrates in (\ref{2.64}) over the multipliers $%
\left( \mu _{(2)}^A,\lambda ^A\right) $, and also over $\left( \Phi
_A^{*(1)},\bar \Phi _A\right) $ one re-obtains (\ref{2.63}) for the choice (%
\ref{2.61}) of $\psi $. One can show by direct computation that the
effective action is both BRST and anti-BRST invariant.
	
\section{Irreducible antifield BRST-anti-BRST quantization of reducible
gauge theories}
\noindent
The basic purpose of this section is to show how reducible gauge theories
can be quantized along the irreducible antifield BRST-anti-BRST formalism.
This task can be accomplished by: (i) building an irreducible theory
equivalent to the original reducible one in a way that allows us to
substitute the BRST quantization of the redundant system by that of the
irreducible one, and (ii) quantizing the resulting irreducible system within
the BRST-anti-BRST framework. The legitimacy of (ii) is implied by (i) and
also by the fact that the BRST-anti-BRST symmetry for a given theory exists
provided the standard BRST symmetry for that theory can be enforced.

\subsection{Irreducible theories associated with reducible ones}
\noindent
Our starting point is a gauge invariant Lagrangian action
\begin{equation}
\label{17}S_0\left[ \Phi ^{\alpha _0}\right] =\int d^Dx{\cal L}_0\left( \Phi
^{\alpha _0},\partial _\mu \Phi ^{\alpha _0},\cdots ,\partial _{\mu
_1}\cdots \partial _{\mu _l}\Phi ^{\alpha _0}\right) , 
\end{equation}
subject to the gauge transformations 
\begin{equation}
\label{18}\delta _\epsilon \Phi ^{\alpha _0}=Z_{\;\;\alpha _1}^{\alpha
_0}\epsilon ^{\alpha _1},\;\alpha _0=1,\cdots ,M_0,\;\alpha _1=1,\cdots
,M_1, 
\end{equation}
which are assumed to be $L$-stage reducible 
\begin{equation}
\label{19}Z_{\;\;\alpha _1}^{\alpha _0}Z_{\;\;\alpha _2}^{\alpha
_1}=C_{\alpha _2}^{\alpha _0\beta _0}\frac{\delta S_0}{\delta \Phi ^{\beta
_0}},\;\alpha _2=1,\cdots ,M_2, 
\end{equation}
\begin{equation}
\label{20}Z_{\;\;\alpha _2}^{\alpha _1}Z_{\;\;\alpha _3}^{\alpha
_2}=C_{\alpha _3}^{\alpha _1\beta _0}\frac{\delta S_0}{\delta \Phi ^{\beta
_0}},\;\alpha _3=1,\cdots ,M_3, 
\end{equation}
$$
\vdots 
$$
\begin{equation}
\label{21}Z_{\;\;\alpha _{L-1}}^{\alpha _{L-2}}Z_{\;\;\alpha _L}^{\alpha
_{L-1}}=C_{\alpha _L}^{\alpha _{L-2}\beta _0}\frac{\delta S_0}{\delta \Phi
^{\beta _0}},\;\alpha _L=1,\cdots ,M_L, 
\end{equation}
\begin{equation}
\label{22}Z_{\;\;\alpha _L}^{\alpha _{L-1}}Z_{\;\;\alpha _{L+1}}^{\alpha
_L}=C_{\alpha _{L+1}}^{\alpha _{L-1}\beta _0}\frac{\delta S_0}{\delta \Phi
^{\beta _0}},\;\alpha _{L+1}=1,\cdots ,M_{L+1}, 
\end{equation}
where $L$ is supposed finite. For the sake of notational simplicity we take
the original fields to be bosonic, but the analysis can be extended to
fermions modulo adding some appropriate sign factors. It is understood that
the functions $Z_{\;\;\alpha _1}^{\alpha _0}$ and $\left( Z_{\;\;\alpha
_{k+1}}^{\alpha _k}\right) _{k=1,\cdots ,L}$ form a complete set of gauge
generators, respectively, reducibility functions.

Initially, we construct an irreducible theory associated with the reducible
one. In this respect, from (\ref{19}--\ref{22}) we remark that 
\begin{equation}
\label{23}{\rm rank}\left( Z_{\;\;\alpha _k}^{\alpha _{k-1}}\right) \approx
\sum\limits_{i=k}^{L+1}\left( -\right) ^{k+i}M_i,\;k=1,\cdots ,L+1, 
\end{equation}
where the weak equality `$\approx $' means an equality valid on the
stationary surface of field equations $\delta S_0/\delta \Phi ^{\alpha _0}=0$%
. Let $\left( A_{\beta _{k-1}}^{\;\;\beta _k}\right) _{k=1,\cdots ,L+1}$ be
some matrices that may involve the fields $\Phi ^{\alpha _0}$, taken such
that 
\begin{equation}
\label{24}{\rm rank}\left( D_{\;\;\alpha _k}^{\beta _k}\right) \approx
\sum\limits_{i=k}^{L+1}\left( -\right) ^{k+i}M_i,\;k=1,\cdots ,L+1, 
\end{equation}
where 
\begin{equation}
\label{25}D_{\;\;\alpha _k}^{\beta _k}=A_{\alpha _{k-1}}^{\;\;\beta
_k}Z_{\;\;\alpha _k}^{\alpha _{k-1}},\;k=1,\cdots ,L+1. 
\end{equation}
In particular we can take $A_{\beta _{k-1}}^{\;\;\beta _k}=\left(
Z_{\;\;\beta _k}^{\beta _{k-1}}\right) ^T$, with $\left( Z_{\;\;\beta
_k}^{\beta _{k-1}}\right) ^T$ the transposed of $Z_{\;\;\beta _k}^{\beta
_{k-1}}$. From (\ref{25}) it follows directly that 
\begin{equation}
\label{52i}D_{\;\;\alpha _k}^{\beta _k}Z_{\;\;\alpha _{k+1}}^{\alpha
_k}\approx 0,\;k=1,\cdots ,L. 
\end{equation}
Relations (\ref{52i}) allow us to represent $D_{\;\;\alpha _k}^{\beta _k}$
under the form 
\begin{equation}
\label{52j}D_{\;\;\alpha _k}^{\beta _k}=\delta _{\;\;\alpha _k}^{\beta
_k}-Z_{\;\;\alpha _{k+1}}^{\beta _k}A_{\alpha _k}^{\;\;\alpha
_{k+1}},\;k=1,\cdots ,L+1. 
\end{equation}
Throughout the paper we work with the conventions 
\begin{equation}
\label{conv}f^{\alpha _k}=0\;{\rm if}\;k<0\;{\rm or}\;k>L+1. 
\end{equation}
It is easy to see that (\ref{52j}) satisfy (\ref{52i}). With these
observations at hand, we pass to the concrete construction of an irreducible
theory associated with the starting reducible one. In view of this, we add
the fields $\Phi ^{\alpha _{2k}}$, $k=1,\ldots ,a$, and the gauge parameters 
$\epsilon ^{\alpha _{2k+1}}$, $k=1,\ldots ,b$, corresponding to every
reducibility relation (\ref{19}--\ref{22}) with even, respectively, odd free
indices, where 
\begin{equation}
\label{53}a=\left\{ 
\begin{array}{c}
\frac L2,\; 
{\rm for}\;L\;{\rm even}, \\ \frac{L+1}2,\;{\rm for}\;L\;{\rm odd}, 
\end{array}
\right. \;b=\left\{ 
\begin{array}{c}
\frac L2,\; 
{\rm for}\;L\;{\rm even}, \\ \frac{L-1}2,\;{\rm for}\;L\;{\rm odd}. 
\end{array}
\right. 
\end{equation}
Under these considerations, we associate the theory described by the action 
\begin{equation}
\label{55}S_0\left[ \Phi ^{\alpha _0},\Phi ^{\alpha _{2k}}\right] =S_0\left[
\Phi ^{\alpha _0}\right] , 
\end{equation}
and subject to the gauge transformations 
\begin{equation}
\label{56}\delta _\epsilon \Phi ^{\alpha _0}=Z_{\;\;\alpha _1}^{\alpha
_0}\epsilon ^{\alpha _1}, 
\end{equation}
\begin{equation}
\label{57}\delta _\epsilon \Phi ^{\alpha _2}=A_{\alpha _1}^{\;\;\alpha
_2}\epsilon ^{\alpha _1}+Z_{\;\;\alpha _3}^{\alpha _2}\epsilon ^{\alpha _3}, 
\end{equation}
$$
\vdots 
$$
\begin{equation}
\label{58}\delta _\epsilon \Phi ^{\alpha _{2k}}=A_{\alpha
_{2k-1}}^{\;\;\alpha _{2k}}\epsilon ^{\alpha _{2k-1}}+Z_{\;\;\alpha
_{2k+1}}^{\alpha _{2k}}\epsilon ^{\alpha _{2k+1}}, 
\end{equation}
$$
\vdots 
$$
\begin{equation}
\label{59}\delta _\epsilon \Phi ^{\alpha _{2a}}=\left\{ 
\begin{array}{l}
A_{\alpha _{L-1}}^{\;\;\alpha _L}\epsilon ^{\alpha _{L-1}}+Z_{\;\;\alpha
_{L+1}}^{\alpha _L}\epsilon ^{\alpha _{L+1}},\; 
{\rm for}\;L\;{\rm even}, \\ A_{\alpha _L}^{\;\;\alpha _{L+1}}\epsilon
^{\alpha _L},\;\;{\rm for}\;L\;{\rm odd}, 
\end{array}
\right. 
\end{equation}
with the starting reducible system. In (\ref{57}--\ref{59}) $A_{\alpha
_{2k-1}}^{\;\;\alpha _{2k}}$ are some matrices that satisfy (\ref{24}). It
is obvious that the transformations (\ref{56}--\ref{59}) leave the action (%
\ref{55}) invariant. From (\ref{55}) we observe that the weak equality
associated with the new system coincides with that corresponding to the
original theory because the field equations of the supplementary fields are
trivial.

Now, it is easy to show that the gauge transformations (\ref{56}--\ref{59})
are irreducible. If we take 
\begin{equation}
\label{61}\epsilon ^{\alpha _{2k+1}}=Z_{\;\;\alpha _{2k+2}}^{\alpha
_{2k+1}}\theta ^{\alpha _{2k+2}}, 
\end{equation}
with arbitrary $\theta $'s, the transformations (\ref{56}--\ref{59}) become 
\begin{equation}
\label{62}\delta _\theta \Phi ^{\alpha _0}\approx 0, 
\end{equation}
\begin{equation}
\label{63}\delta _\theta \Phi ^{\alpha _2}\approx D_{\;\;\beta _2}^{\alpha
_2}\theta ^{\beta _2}, 
\end{equation}
$$
\vdots 
$$
\begin{equation}
\label{64}\delta _\theta \Phi ^{\alpha _{2k}}\approx D_{\;\;\beta
_{2k}}^{\alpha _{2k}}\theta ^{\beta _{2k}}, 
\end{equation}
$$
\vdots 
$$
\begin{equation}
\label{65}\delta _\theta \Phi ^{\alpha _{2a}}\approx \left\{ 
\begin{array}{l}
D_{\;\;\beta _L}^{\alpha _L}\theta ^{\beta _L},\; 
{\rm for}\;L\;{\rm even}, \\ D_{\;\;\beta _{L+1}}^{\alpha _{L+1}}\theta
^{\beta _{L+1}},\;\;{\rm for}\;L\;{\rm odd}. 
\end{array}
\right. 
\end{equation}
Using (\ref{52i}) and the completeness of the reducibility functions, it
follows that $\delta _\theta \Phi ^{\alpha _{2k}}\approx 0$, $k=0,\cdots ,a$%
, if and only if 
\begin{equation}
\label{66}\theta ^{\beta _{2k}}\approx Z_{\;\;\beta _{2k+1}}^{\beta
_{2k}}\lambda ^{\beta _{2k+1}},\;k=1,\cdots ,a, 
\end{equation}
for some arbitrary functions $\lambda ^{\beta _{2k+1}}$. Inserting (\ref{66}%
) in (\ref{61}), it results that 
\begin{equation}
\label{if}\delta _\epsilon \Phi ^{\alpha _{2k}}\approx 0\Leftrightarrow
\epsilon ^{\alpha _{2k+1}}\approx 0, 
\end{equation}
so the gauge transformations with the parameters (\ref{61}) are trivial.
This establishes the irreducibility of the gauge transformations (\ref{56}--%
\ref{59}). The prior construction of the irreducible gauge transformations
does not guarantee their completeness. In fact, it is impossible to proof in
general the completeness of the irreducible gauge transformations (\ref{56}--%
\ref{59}) as it depends on the choice of the matrices $A_{\alpha
_{k-1}}^{\;\;\alpha _k}$ and also on the original reducibility matrices.
This is why in the sequel we assume the completeness of (\ref{56}--\ref{59}%
). This feature is required by the possibility to construct a weakly
nilpotent longitudinal exterior derivative along the gauge orbits connected
with the irreducible theory.

At this point we investigate whether it is legitimate or not to replace the
BRST quantization of the reducible theory by that of the irreducible system
constructed previously. Initially, we construct the BRST symmetry associated
with the irreducible system. First, we derive the Koszul-Tate differential
in a way that ensures its acyclicity. The minimal antifield spectrum
includes the fermionic antifields 
\begin{equation}
\label{79}\left( \Phi _{\alpha _0}^{*},\Phi _{\alpha _{2k}}^{*}\right)
,\;k=1,\cdots ,a, 
\end{equation}
with antighost number one, and the bosonic antifields 
\begin{equation}
\label{80}\eta _{\alpha _{2k+1}}^{*},\;k=0,\cdots ,b, 
\end{equation}
with antighost number two. The irreducible Koszul-Tate operator, $\delta $,
acts on its generators through 
\begin{equation}
\label{81}\delta \Phi ^{\alpha _0}=0,\;\delta \Phi ^{\alpha
_{2k}}=0,\;k=1,\cdots ,a, 
\end{equation}
\begin{equation}
\label{82}\delta \Phi _{\alpha _0}^{*}=-\frac{\delta S_0}{\delta \Phi
^{\alpha _0}}, 
\end{equation}
\begin{equation}
\label{83}\delta \Phi _{\alpha _{2k}}^{*}=-\frac{\delta S_0}{\delta \Phi
^{\alpha _{2k}}}\equiv 0,\;k=1,\cdots ,a, 
\end{equation}
\begin{equation}
\label{84}\delta \eta _{\alpha _{2k+1}}^{*}=\Phi _{\alpha
_{2k}}^{*}Z_{\;\;\alpha _{2k+1}}^{\alpha _{2k}}+\Phi _{\alpha
_{2k+2}}^{*}A_{\alpha _{2k+1}}^{\;\;\alpha _{2k+2}},\;k=0,\cdots ,b, 
\end{equation}
such that $\delta $ is clearly nilpotent, $\delta ^2=0$. Relations (\ref{82}%
--\ref{83}) give rise to the co-cycles $\Phi _{\alpha _0}^{*}Z_{\;\;\alpha
_1}^{\alpha _0}$ and $\Phi _{\alpha _{2k}}^{*}$, with $k=1,\cdots ,a$, while
(\ref{84}) lead to some combinations of these co-cycles which are $\delta $%
-exact. However, this does not ensure the separate $\delta $-exactness of $%
\Phi _{\alpha _0}^{*}Z_{\;\;\alpha _1}^{\alpha _0}$ and $\Phi _{\alpha
_{2k}}^{*}$. Let us prove that these co-cycles are indeed $\delta $-exact.
For definiteness we expose the case $L$ even, the other situation being
solved in the same fashion. Multiplying (\ref{84}) for $k=L/2-1$ by $%
Z_{\;\;\beta _L}^{\alpha _{L-1}}$ and using (\ref{52j}) we find 
\begin{equation}
\label{85a}\Phi _{\beta _L}^{*}=\delta \gamma _{\beta _L}, 
\end{equation}
with 
\begin{equation}
\label{85b}\gamma _{\beta _L}=\left( \eta _{\alpha _{L-1}}^{*}Z_{\;\;\beta
_L}^{\alpha _{L-1}}+\alpha _2\Phi _{\alpha _{L-2}}^{*}\Phi _{\alpha
_0}^{*}C_{\;\;\beta _L}^{\alpha _{L-2}\alpha _0}+\eta _{\alpha
_{L+1}}^{*}A_{\beta _L}^{\;\;\alpha _{L+1}}\right) , 
\end{equation}
where $\alpha _2=1/2$ for $L=2$, and $\alpha _2=1$ otherwise. Then, (\ref{84}%
) for $k=L/2-1$ takes the form 
\begin{equation}
\label{86}\delta \left( \eta _{\alpha _{L-1}}^{*}-A_{\alpha
_{L-1}}^{\;\;\beta _L}\gamma _{\beta _L}\right) =\Phi _{\alpha
_{L-2}}^{*}Z_{\;\;\alpha _{L-1}}^{\alpha _{L-2}}. 
\end{equation}
Now, we multiply the relation corresponding to (\ref{84}) for $k=L/2-2$ by $%
Z_{\;\;\beta _{L-2}}^{\alpha _{L-3}}$, and consequently obtain 
\begin{equation}
\label{87}\delta \left( \eta _{\alpha _{L-3}}^{*}Z_{\;\;\beta
_{L-2}}^{\alpha _{L-3}}+\alpha _4\Phi _{\alpha _{L-4}}^{*}\Phi _{\alpha
_0}^{*}C_{\;\;\beta _{L-2}}^{\alpha _{L-4}\alpha _0}\right) =\Phi _{\alpha
_{L-2}}^{*}D_{\;\;\beta _{L-2}}^{\alpha _{L-2}}, 
\end{equation}
where $\alpha _4=1/2$ for $L=4$, and $\alpha _4=1$ otherwise. Replacing (\ref
{52j}) in (\ref{87}) and employing (\ref{86}) it results that $\Phi _{\beta
_{L-2}}^{*}$ is also $\delta $-exact. Reprising the same procedure for each
level we infer 
\begin{equation}
\label{88}\Phi _{\alpha _0}^{*}Z_{\;\;\alpha _1}^{\alpha _0}=\delta \gamma
_{\alpha _1},\;\Phi _{\beta _{2k}}^{*}=\delta \gamma _{\beta
_{2k}},\;k=1,\cdots ,a, 
\end{equation}
with 
\begin{eqnarray}\label{89a}
& &\gamma _{\beta _{2k}}= 
\left( Z_{\;\;\beta _{2k}}^{\beta _{2k-1}}
\eta _{\beta _{2k-1}}^{*}+\alpha _{L-2k+2}
C_{\;\;\beta _{2k}}^{\beta _{2k-2}\alpha _0}
\Phi _{\beta _{2k-2}}^{*}
\Phi _{\alpha _0}^{*}+\right. \nonumber \\
& &\left. A_{\beta _{2k}}^{\;\;\beta _{2k+1}}
\left( \eta _{\beta _{2k+1}}^{*}-
A_{\beta _{2k+1}}^{\;\;\beta _{2k+2}}
\gamma _{\beta _{2k+2}}\right) \right) , 
\end{eqnarray}
\begin{equation}
\label{89b}\gamma _{\alpha _1}=\eta _{\alpha _1}^{*}-A_{\alpha
_1}^{\;\;\alpha _2}\gamma _{\alpha _2}, 
\end{equation}
and $\alpha _{L-2k+2}=1$ for $L\neq 2k-2$, respectively, $\alpha
_{L-2k+2}=1/2$ for $L=2k-2$. The last relations enforce the triviality of
the above mentioned co-cycles at antighost number one. Moreover, there are
no non trivial cocycles at resolution degrees greater than one due to the
irreducibility of the gauge transformations (\ref{56}--\ref{59}), hence the
irreducible Koszul-Tate differential is acyclic.

The construction of the longitudinal exterior differential along the gauge
orbits, $d$, follows the general irreducible BRST line \cite{5}, the
hypothesis of completeness on the irreducible gauge transformations ensuring
the weak nilpotency of $d$ without introducing any ghosts of ghosts. Under
these considerations, the homological perturbation theory \cite{31}--\cite
{34} guarantees the existence of the irreducible BRST symmetry, $s_I$. We
observe that the two theories display the same classical observables as the
fields $\left( \Phi ^{\alpha _{2k}}\right) _{k=1,\ldots ,a}$ are not
effectively involved with the action (\ref{55}) (of the irreducible system),
being therefore purely gauge. Consequently, the observables of the
irreducible theory do not depend on these fields and check the equations 
\begin{equation}
\label{obs}\frac{\delta F}{\delta \Phi ^{\alpha _0}}Z_{\;\;\alpha
_1}^{\alpha _0}\approx 0, 
\end{equation}
which are nothing but the equations that must be verified by the observables
of the reducible theory. As the observables of the irreducible and reducible
theories coincide, the zeroth order cohomological groups corresponding to
the irreducible and reducible formulations are also equal 
\begin{equation}
\label{h}H^0\left( s_I\right) =H^0\left( s_R\right) , 
\end{equation}
with $s_R$ denoting the reducible BRST symmetry. Hence, the irreducible and
reducible theories are equivalent from the BRST point of view, i.e., from
the point of view of the basic equations describing this formalism 
\begin{equation}
\label{eq}s^2=0,\;H^0\left( s\right) =\left\{ {\rm physical\;observables}%
\right\} . 
\end{equation}
The last conclusion ensures that we can substitute the BRST quantization of
the reducible theory by that of the irreducible system derived at the
beginning of this section. Taking into account that both the existence and
construction of the antifield BRST-anti-BRST symmetry are essentially based
on the existence of the standard Lagrangian BRST symmetry, it follows that
we can safely replace the BRST-anti-BRST quantization of the original
reducible theory with that of the irreducible system. This is the concern of
the next subsection.

\subsection{Irreducible antifield BRST-anti-BRST quantization}
\noindent
In this subsection we perform the BRST-anti-BRST quantization of the
irreducible theory constructed in the previous subsection, which is
described by the action (\ref{55}) and is subject to the irreducible gauge
transformations (\ref{56}--\ref{59}). The initial field spectrum and gauge
parameters are respectively given by $\left( \Phi ^{\alpha _{2k}}\right)
_{k=0,\cdots ,a}$ and $\left( \epsilon ^{\alpha _{2k+1}}\right) _{k=0,\cdots
,b}$, hence we can make the following analogies between (\ref{2.2}) and our
system 
\begin{equation}
\label{4.1}\Phi ^i\leftrightarrow \left( \Phi ^{\alpha _{2k}}\right)
_{k=0,\cdots ,a},\;\epsilon ^\alpha \leftrightarrow \left( \epsilon ^{\alpha
_{2k+1}}\right) _{k=0,\cdots ,b}, 
\end{equation}
where $R_{\;\;\alpha }^i$ is expressed in our case by a matrix containing $%
(a+1)\times (b+1)$ blocks of elements $Z_{\;\;\alpha _{2k+1}}^{\alpha _{2k}}$
and $A_{\alpha _{2k-1}}^{\;\;\alpha _{2k}}$ structured accordingly (\ref{56}%
--\ref{59}). In agreement with the discussion from Section 2, the field,
ghost and antifield spectra (see (\ref{2.11}) and (\ref{2.26}--\ref{2.28}))
are organized as 
\begin{equation}
\label{4.2}\Phi ^A=\left( \stackrel{\left( 0,0\right) }{\Phi }^{\alpha
_{2k}},\stackrel{\left( 1,0\right) }{\eta }_1^{\alpha _{2k+1}},\stackrel{%
\left( 0,1\right) }{\eta }_2^{\alpha _{2k+1}},\stackrel{\left( 1,1\right) }{%
\pi }^{\alpha _{2k+1}}\right) , 
\end{equation}
\begin{equation}
\label{4.3}\Phi _A^{*(1)}=\left( \stackrel{(-1,0)}{\Phi }_{\alpha
_{2k}}^{*(1)},\stackrel{(-2,0)}{\eta }_{\alpha _{2k+1}}^{*(11)},\stackrel{%
(-1,-1)}{\eta }_{\alpha _{2k+1}}^{*(12)},\stackrel{(-2,-1)}{\pi }_{\alpha
_{2k+1}}^{*(1)}\right) , 
\end{equation}
\begin{equation}
\label{4.4}\Phi _A^{*(2)}=\left( \stackrel{(0,-1)}{\Phi }_{\alpha
_{2k}}^{*(2)},\stackrel{(-1,-1)}{\eta }_{\alpha _{2k+1}}^{*(21)},\stackrel{%
(0,-2)}{\eta }_{\alpha _{2k+1}}^{*(22)},\stackrel{(-1,-2)}{\pi }_{\alpha
_{2k+1}}^{*(2)}\right) , 
\end{equation}
\begin{equation}
\label{4.5}\bar \Phi _A=\left( \stackrel{(-1,-1)}{\bar \Phi }_{\alpha _{2k}},%
\stackrel{(-2,-1)}{\bar \eta }_{1\alpha _{2k+1}},\stackrel{(-1,-2)}{\bar
\eta }_{2\alpha _{2k+1}},\stackrel{(-2,-2)}{\bar \pi }_{\alpha
_{2k+1}}\right) , 
\end{equation}
where the superscript indicates the bighost number.

The definitions of $\delta _1$ and $\delta _2$ (see (\ref{2.28a}--\ref{2.33}%
)) on the generators from the BRST-anti-BRST complex are expressed by 
\begin{equation}
\label{4.5a}\delta _1\stackrel{(0,0)}{\Phi }^A=\delta _2\stackrel{(0,0)}{%
\Phi }^A=0,
\end{equation}
\begin{equation}
\label{4.6}\delta _1\stackrel{(-1,0)}{\Phi }_{\alpha _0}^{*(1)}=\delta _2%
\stackrel{(0,-1)}{\Phi }_{\alpha _0}^{*(2)}=-\frac{\delta S_0}{\delta \Phi
^{\alpha _0}},
\end{equation}
\begin{equation}
\label{4.8}\delta _1\stackrel{(-1,0)}{\Phi }_{\alpha _{2k}}^{*(1)}=\delta _2%
\stackrel{(0,-1)}{\Phi }_{\alpha _{2k}}^{*(2)}=-\frac{\delta S_0}{\delta
\Phi ^{\alpha _{2k}}}\equiv 0,\;k\geq 1,
\end{equation}
\begin{equation}
\label{4.9}\delta _1\stackrel{(-2,0)}{\eta }_{\alpha _{2k+1}}^{*(11)}=\delta
_2\stackrel{(-1,-1)}{\eta }_{\alpha _{2k+1}}^{*(21)}=\stackrel{(-1,0)}{\Phi }%
_{\alpha _{2k}}^{*(1)}Z_{\;\;\alpha _{2k+1}}^{\alpha _{2k}}+\stackrel{(-1,0)%
}{\Phi }_{\alpha _{2k+2}}^{*(1)}A_{\alpha _{2k+1}}^{\;\;\alpha
_{2k+2}},k\geq 0,
\end{equation}
\begin{equation}
\label{4.10}\delta _1\stackrel{(-1,-1)}{\eta }_{\alpha
_{2k+1}}^{*(12)}=\delta _2\stackrel{(0,-2)}{\eta }_{\alpha _{2k+1}}^{*(22)}=%
\stackrel{(0,-1)}{\Phi }_{\alpha _{2k}}^{*(2)}Z_{\;\;\alpha _{2k+1}}^{\alpha
_{2k}}+\stackrel{(0,-1)}{\Phi }_{\alpha _{2k+2}}^{*(2)}A_{\alpha
_{2k+1}}^{\;\;\alpha _{2k+2}},k\geq 0,
\end{equation}
\begin{eqnarray}\label{4.11}
& &\delta _1\stackrel{(-2,-1)}{\pi }_{\alpha _{2k+1}}^{*(1)}=
\delta _2\stackrel{(-1,-2)}{\pi }_{\alpha _{2k+1}}^{*(2)}=
\stackrel{(-1,-1)}{\eta }_{\alpha
_{2k+1}}^{*(12)}-\stackrel{(-1,-1)}{\eta }_{\alpha _{2k+1}}^{*(21)}- 
\nonumber \\
& &\stackrel{(-1,-1)}{\bar \Phi }_{\alpha _{2k}}Z_{\;\;\alpha
_{2k+1}}^{\alpha _{2k}}-\stackrel{(-1,-1)}{\bar \Phi }_{\alpha
_{2k+2}}A_{\alpha _{2k+1}}^{\;\;\alpha _{2k+2}},\;k\geq 0, 
\end{eqnarray}
\begin{equation}
\label{4.12}\delta _1\Phi _A^{*(2)}=0=\delta _2\Phi _A^{*(1)},
\end{equation}
\begin{equation}
\label{4.13}\delta _1\stackrel{(-1,-1)}{\bar \Phi }_{\alpha _{2k}}=\stackrel{%
(0,-1)}{\Phi }_{\alpha _{2k}}^{*(2)},\;\delta _2\stackrel{(-1,-1)}{\bar \Phi 
}_{\alpha _{2k}}=-\stackrel{(-1,0)}{\Phi }_{\alpha _{2k}}^{*(1)},\;k\geq 0,
\end{equation}
\begin{equation}
\label{4.14}\delta _1\stackrel{(-2,-1)}{\bar \eta }_{1\alpha _{2k+1}}=%
\stackrel{(-1,-1)}{\eta }_{\alpha _{2k+1}}^{*(21)},\;\delta _2\stackrel{%
(-2,-1)}{\bar \eta }_{1\alpha _{2k+1}}=-\stackrel{(-2,0)}{\eta }_{\alpha
_{2k+1}}^{*(11)},\;k\geq 0,
\end{equation}
\begin{equation}
\label{4.15}\delta _1\stackrel{(-1,-2)}{\bar \eta }_{2\alpha _{2k+1}}=%
\stackrel{(0,-2)}{\eta }_{\alpha _{2k+1}}^{*(22)},\;\delta _2\stackrel{%
(-1,-2)}{\bar \eta }_{2\alpha _{2k+1}}=-\stackrel{(-1,-1)}{\eta }_{\alpha
_{2k+1}}^{*(12)},\;k\geq 0,
\end{equation}
\begin{equation}
\label{4.16}\delta _1\stackrel{(-2,-2)}{\bar \pi }_{\alpha _{2k+1}}=%
\stackrel{(-1,-2)}{\pi }_{\alpha _{2k+1}}^{*(2)},\;\delta _2\stackrel{(-2,-2)%
}{\bar \pi }_{\alpha _{2k+1}}=-\stackrel{(-2,-1)}{\pi }_{\alpha
_{2k+1}}^{*(1)},\;k\geq 0.
\end{equation}
As we have already discussed in the previous subsection, there appear some
problems connected with the existence of some apparently non trivial
co-cycles for the Koszul-Tate operator in the context of the antifield BRST
formulation (see formulas (\ref{82}--\ref{83})). The same problem is present
within the BRST-anti-BRST approach, namely, the co-cycles $\stackrel{(-1,0)}{%
\Phi }_{\alpha _0}^{*(1)}Z_{\;\;\alpha _1}^{\alpha _0}$, $\left( \stackrel{%
(-1,0)}{\Phi }_{\alpha _{2k}}^{*(1)}\right) _{k\geq 1}$ and $\stackrel{(0,-1)%
}{\Phi }_{\alpha _0}^{*(2)}Z_{\;\;\alpha _1}^{\alpha _0}$, $\left( \stackrel{%
(0,-1)}{\Phi }_{\alpha _{2k}}^{*(2)}\right) _{k\geq 1}$ (obtained from (\ref
{4.6}) multiplied by $Z_{\;\;\alpha _1}^{\alpha _0}$ and (\ref{4.8})) are
both $\delta _1$- and $\delta _2$-closed. However, the first two sets of
co-cycles are killed in the homology of $\delta _2$, and the same is for the
other two sets, but in the homology of $\delta _1$ (see (\ref{4.13})). In
fact, all the co-cycles $\Phi _A^{*(2)}$ and $\Phi _A^{*(1)}$ are dropped
out from the homology of $\delta _1$, respectively, $\delta _2$ by means of
the bar variables (see \ref{4.13}--\ref{4.16}). In consequence, the only
dangerous co-cycles are represented by $\stackrel{(-1,0)}{\Phi }_{\alpha
_0}^{*(1)}Z_{\;\;\alpha _1}^{\alpha _0}$, $\left( \stackrel{(-1,0)}{\Phi }%
_{\alpha _{2k}}^{*(1)}\right) _{k\geq 1}$ in the homology of $\delta _1$,
and $\stackrel{(0,-1)}{\Phi }_{\alpha _0}^{*(2)}Z_{\;\;\alpha _1}^{\alpha _0}
$, $\left( \stackrel{(0,-1)}{\Phi }_{\alpha _{2k}}^{*(2)}\right) _{k\geq 1}$
in the homology of $\delta _2$. In order to perform a proper construction of
the Koszul-Tate bicomplex it is necessary to investigate their exactness.
The proof showing the $\delta _1$-, respectively, $\delta _2$-exactness of
the above invoked co-cycles will be done below for definiteness in the case $%
L$ even, but the opposite situation can be solved in a similar fashion.

We start from the last two relations (\ref{4.9}) with respect to $\delta _1$
in the hypothesis $L$ even 
\begin{equation}
\label{4.17}\delta _1\stackrel{(-2,0)}{\eta }_{\alpha _{L-1}}^{*(11)}=%
\stackrel{(-1,0)}{\Phi }_{\alpha _{L-2}}^{*(1)}Z_{\;\;\alpha _{L-1}}^{\alpha
_{L-2}}+\stackrel{(-1,0)}{\Phi }_{\alpha _L}^{*(1)}A_{\alpha
_{L-1}}^{\;\;\alpha _L}, 
\end{equation}
\begin{equation}
\label{4.18}\delta _1\stackrel{(-2,0)}{\eta }_{\alpha _{L+1}}^{*(11)}=%
\stackrel{(-1,0)}{\Phi }_{\alpha _L}^{*(1)}Z_{\;\;\alpha _{L+1}}^{\alpha
_L}. 
\end{equation}
If we multiply (\ref{4.18}) by $A_{\beta _L}^{\;\;\alpha _{L+1}}$, (\ref
{4.17}) by $Z_{\;\;\beta _L}^{\alpha _{L-1}}$ and sum the resulting
relations, we find%
\begin{eqnarray}\label{4.19}
& &\delta _1\left( \stackrel{(-2,0)}{\eta }_{\alpha _{L-1}}^{*(11)}%
Z_{\;\;\beta
_L}^{\alpha _{L-1}}+\stackrel{(-2,0)}{\eta }_{\alpha _{L+1}}^{*(11)}A_{\beta
_L}^{\;\;\alpha _{L+1}}\right) =\stackrel{(-1,0)}{\Phi }_{\alpha
_{L-2}}^{*(1)}Z_{\;\;\alpha _{L-1}}^{\alpha _{L-2}}Z_{\;\;\beta _L}^{\alpha
_{L-1}}+\nonumber \\ 
& &\stackrel{(-1,0)}{\Phi }_{\alpha _L}^{*(1)}\left( A_{\alpha
_{L-1}}^{\;\;\alpha _L}Z_{\;\;\beta _L}^{\alpha _{L-1}}+Z_{\;\;\alpha
_{L+1}}^{\alpha _L}A_{\beta _L}^{\;\;\alpha _{L+1}}\right) . 
\end{eqnarray}
Taking now into account (\ref{21}), (\ref{52j}) for $k=L$, (\ref{4.6}) with
respect to $\delta _1$ and also (\ref{4.8}) for $k=\frac L2-1$ (and with
respect to $\delta _1$), it follows that 
\begin{equation}
\label{4.20}\stackrel{(-1,0)}{\Phi }_{\beta _L}^{*(1)}=\delta _1\stackrel{%
(-2,0)}{\gamma }_{\beta _L}, 
\end{equation}
where 
\begin{eqnarray}\label{4.21}
& &\stackrel{(-2,0)}{\gamma }_{\beta _L}=\stackrel{(-2,0)}{\eta }_{\alpha
_{L-1}}^{*(11)}Z_{\;\;\beta _L}^{\alpha _{L-1}}+\nonumber \\ 
& &a_2\stackrel{(-1,0)}{\Phi }_{\alpha _{L-2}}^{*(1)}C_{\beta
_L}^{\alpha _{L-2}\beta _0}\stackrel{(-1,0)}{\Phi }_{\beta _0}^{*(1)}+%
\stackrel{(-2,0)}{\eta }_{\alpha _{L+1}}^{*(11)}A_{\beta _L}^{\;\;\alpha
_{L+1}}, 
\end{eqnarray}
with 
\begin{equation}
\label{4.22}a_2=\left\{ 
\begin{array}{c}
1,\; 
{\rm if}\;L\neq 2, \\ \frac 12,\;{\rm if}\;L=2. 
\end{array}
\right. 
\end{equation}
Inserting the relations (\ref{4.20}) in (\ref{4.17}), we have that 
\begin{equation}
\label{4.23}\stackrel{(-1,0)}{\Phi }_{\alpha _{L-2}}^{*(1)}Z_{\;\;\alpha
_{L-1}}^{\alpha _{L-2}}=\delta _1\left( \stackrel{(-2,0)}{\eta }_{\alpha
_{L-1}}^{*(11)}-\stackrel{(-2,0)}{\gamma }_{\alpha _L}A_{\alpha
_{L-1}}^{\;\;\alpha _L}\right) . 
\end{equation}
Next, we pass to the definitions (\ref{4.9}) for $k=\frac L2-2$ (with
respect to $\delta _1$) 
\begin{equation}
\label{4.24}\delta _1\stackrel{(-2,0)}{\eta }_{\alpha _{L-3}}^{*(11)}=%
\stackrel{(-1,0)}{\Phi }_{\alpha _{L-4}}^{*(1)}Z_{\;\;\alpha _{L-3}}^{\alpha
_{L-4}}+\stackrel{(-1,0)}{\Phi }_{\alpha _{L-2}}^{*(1)}A_{\alpha
_{L-3}}^{\;\;\alpha _{L-2}}, 
\end{equation}
and multiply these relations by $Z_{\;\;\beta _{L-2}}^{\alpha _{L-3}}$,
arriving at 
\begin{equation}
\label{4.25}\delta _1\left( \stackrel{(-2,0)}{\eta }_{\alpha
_{L-3}}^{*(11)}Z_{\;\;\beta _{L-2}}^{\alpha _{L-3}}\right) =\stackrel{(-1,0)%
}{\Phi }_{\alpha _{L-4}}^{*(1)}Z_{\;\;\alpha _{L-3}}^{\alpha
_{L-4}}Z_{\;\;\beta _{L-2}}^{\alpha _{L-3}}+\stackrel{(-1,0)}{\Phi }_{\alpha
_{L-2}}^{*(1)}A_{\alpha _{L-3}}^{\;\;\alpha _{L-2}}Z_{\;\;\beta
_{L-2}}^{\alpha _{L-3}}. 
\end{equation}
If we take into consideration the reducibility relations (\ref{19}--\ref{22}%
) for $Z_{\;\;\alpha _{L-3}}^{\alpha _{L-4}}Z_{\;\;\beta _{L-2}}^{\alpha
_{L-3}}$, (\ref{4.6}) with respect to $\delta _1$ and (\ref{52j}) for $k=L-2$%
, we are led to%
\begin{eqnarray}\label{4.26}
& &\delta _1\left( \stackrel{(-2,0)}{\eta }_{\alpha _{L-3}}^{*(11)}Z_{\;\;\beta
_{L-2}}^{\alpha _{L-3}}+a_4\stackrel{(-1,0)}{\Phi }_{\alpha
_{L-4}}^{*(1)}C_{\beta _{L-2}}^{\alpha _{L-4}\beta _0}\stackrel{(-1,0)}{\Phi 
}_{\beta _0}^{*(1)}\right) =\nonumber \\ 
& &\stackrel{(-1,0)}{\Phi }_{\beta _{L-2}}^{*(1)}-
\stackrel{(-1,0)}{\Phi }_{\alpha _{L-2}}^{*(1)}
Z_{\;\;\alpha _{L-1}}^{\alpha _{L-2}}A_{\beta
_{L-2}}^{\;\;\alpha _{L-1}}, 
\end{eqnarray}
where 
\begin{equation}
\label{4.27}a_4=\left\{ 
\begin{array}{c}
1,\; 
{\rm if}\;L\neq 4, \\ \frac 12,\;{\rm if}\;L=4. 
\end{array}
\right. 
\end{equation}
At this moment we employ (\ref{4.23}) and derive 
\begin{equation}
\label{4.28}\stackrel{(-1,0)}{\Phi }_{\beta _{L-2}}^{*(1)}=\delta _1%
\stackrel{(-2,0)}{\gamma }_{\beta _{L-2}}, 
\end{equation}
with%
\begin{eqnarray}\label{4.29}
& &\stackrel{(-2,0)}{\gamma }_{\beta _{L-2}}=\stackrel{(-2,0)}{\eta }_{\alpha
_{L-3}}^{*(11)}Z_{\;\;\beta _{L-2}}^{\alpha _{L-3}}+
a_4\stackrel{(-1,0)}{\Phi }_{\alpha _{L-4}}^{*(1)}%
C_{\beta _{L-2}}^{\alpha _{L-4}\beta _0}%
\stackrel{(-1,0)}{\Phi }_{\beta _0}^{*(1)}+\nonumber \\ 
& &\left( \stackrel{(-2,0)}{\eta }_{\alpha _{L-1}}^{*(11)}-%
\stackrel{(-2,0)}{\gamma }_{\alpha _L}A_{\alpha _{L-1}}^{\;\;\alpha
_L}\right) A_{\beta _{L-2}}^{\;\;\alpha _{L-1}}. 
\end{eqnarray}
Substituting (\ref{4.28}) in (\ref{4.24}) we additionally infer 
\begin{equation}
\label{4.30}\stackrel{(-1,0)}{\Phi }_{\alpha _{L-4}}^{*(1)}Z_{\;\;\alpha
_{L-3}}^{\alpha _{L-4}}=\delta _1\left( \stackrel{(-2,0)}{\eta }_{\alpha
_{L-3}}^{*(11)}-\stackrel{(-2,0)}{\gamma }_{\beta _{L-2}}A_{\alpha
_{L-3}}^{\;\;\alpha _{L-2}}\right) . 
\end{equation}
Reprising the same treatment on the other definitions (\ref{4.9}) with
respect to $\delta _1$ we consequently deduce that all the antifields $%
\left( \stackrel{(-1,0)}{\Phi }_{\beta _{L-2k}}^{*(1)}\right) _{k=0,\cdots
,a-1\equiv \frac L2-1}$ are $\delta _1$-exact, i.e., 
\begin{equation}
\label{4.31}\stackrel{(-1,0)}{\Phi }_{\beta _{L-2k}}^{*(1)}=\delta _1%
\stackrel{(-2,0)}{\gamma }_{\beta _{L-2k}}, 
\end{equation}
with%
\begin{eqnarray}\label{4.32}
& &\stackrel{\left( -2,0\right) }{\gamma }_{\beta _{L-2k}}=\stackrel{\left(
-2,0\right) }{\eta }_{\alpha _{L-2k-1}}^{*\left( 11\right) }Z_{\;\;\beta
_{L-2k}}^{\alpha _{L-2k-1}}+a_{2k+2}\stackrel{\left( -1,0\right) }{\Phi }%
_{\alpha _{L-2k-2}}^{*\left( 1\right) }C_{\beta _{L-2k}}^{\alpha
_{L-2k-2}\beta _0}\stackrel{\left( -1,0\right) }{\Phi }_{\beta _0}^{*\left(
1\right) }\nonumber \\ 
& &+\left( \stackrel{(-2,0)}{\eta }_{\alpha _{L-2k+1}}^{*(11)}-%
\stackrel{(-2,0)}{\gamma }_{\alpha _{L-2k+2}}A_{\alpha
_{L-2k+1}}^{\;\;\alpha _{L-2k+2}}\right) A_{\beta _{L-2k}}^{\;\;\alpha
_{L-2k+1}}, 
\end{eqnarray}
where 
\begin{equation}
\label{4.33}a_{2k+2}=\left\{ 
\begin{array}{c}
1,\;L\neq 2k+2, \\ 
\frac 12,\;L=2k+2. 
\end{array}
\right. 
\end{equation}
The definition of $\delta _1$ acting on $\stackrel{(-2,0)}{\eta }_{\alpha
_1}^{*(11)}$%
\begin{equation}
\label{4.34}\delta _1\stackrel{(-2,0)}{\eta }_{\alpha _1}^{*(11)}=\stackrel{%
(-1,0)}{\Phi }_{\alpha _0}^{*(1)}Z_{\;\;\alpha _1}^{\alpha _0}+\stackrel{%
(-1,0)}{\Phi }_{\alpha _2}^{*(1)}A_{\alpha _1}^{\;\;\alpha _2}, 
\end{equation}
together with (\ref{4.31}) for $k=\frac L2-1$ yield 
\begin{equation}
\label{4.35}\stackrel{(-1,0)}{\Phi }_{\alpha _0}^{*(1)}Z_{\;\;\alpha
_1}^{\alpha _0}=\delta _1\left( \stackrel{(-2,0)}{\eta }_{\alpha _1}^{*(11)}-%
\stackrel{(-2,0)}{\gamma }_{\alpha _2}A_{\alpha _1}^{\;\;\alpha _2}\right) . 
\end{equation}
In this way we managed to show that all the dangerous $\delta _1$-co-cycles $%
\left( \stackrel{(-1,0)}{\Phi }_{\alpha _{2k}}^{*(1)}\right) _{k\geq 1}$ and 
$\stackrel{(-1,0)}{\Phi }_{\alpha _0}^{*(1)}Z_{\;\;\alpha _1}^{\alpha _0}$
are indeed $\delta _1$-exact. Following a similar line we can prove that the 
$\delta _2$ co-cycles $\left( \stackrel{(0,-1)}{\Phi }_{\alpha
_{2k}}^{*(2)}\right) _{k\geq 1}$ and $\stackrel{(0,-1)}{\Phi }_{\alpha
_0}^{*(2)}Z_{\;\;\alpha _1}^{\alpha _0}$ are $\delta _2$-exact 
\begin{equation}
\label{4.36}\stackrel{(0,-1)}{\Phi }_{\alpha _{L-2k}}^{*(2)}=\delta _2%
\stackrel{(0,-2)}{\gamma ^{\prime }}_{\alpha _{L-2k}},\;k=0,\cdots
,a-1\equiv \frac L2-1, 
\end{equation}
\begin{equation}
\label{4.37}\stackrel{(0,-1)}{\Phi }_{\alpha _0}^{*(2)}Z_{\;\;\alpha
_1}^{\alpha _0}=\delta _1\left( \stackrel{(0,-2)}{\eta }_{\alpha _1}^{*(22)}-%
\stackrel{(0,-2)}{\gamma ^{\prime }}_{\alpha _2}A_{\alpha _1}^{\;\;\alpha
_2}\right) , 
\end{equation}
where%
\begin{eqnarray}\label{4.38}
& &\stackrel{(0,-2)}{\gamma ^{\prime }}_{\beta _{L-2k}}=
\stackrel{(0,-2)}{\eta }%
_{\alpha _{L-2k-1}}^{*(22)}Z_{\;\;\beta _{L-2k}}^{\alpha _{L-2k-1}}+a_{2k+2}%
\stackrel{(0,-1)}{\Phi }_{\alpha _{L-2k-2}}^{*(2)}C_{\beta _{L-2k}}^{\alpha
_{L-2k-2}\beta _0}\stackrel{(0,-1)}{\Phi }_{\beta _0}^{*(2)} \nonumber \\ 
& &+\left( \stackrel{(0,-2)}{\eta }_{\alpha _{L-2k+1}}^{*(22)}-%
\stackrel{(0,-2)}{\gamma ^{\prime }}_{\alpha _{L-2k+2}}A_{\alpha
_{L-2k+1}}^{\;\;\alpha _{L-2k+2}}\right) A_{\beta _{L-2k}}^{\;\;\alpha
_{L-2k+1}}. 
\end{eqnarray}
In consequence, there are no non trivial co-cycles of $\delta _1$ and $%
\delta _2$ at positive resolution bidegrees because on the one hand we
established that all the co-cycles at resolution bidegree $(1,0)$ or $(0,1)$
are trivial, and, on the other hand, the appearance of non trivial co-cycles
at higher order resolution bidegrees is prevented precisely by the
irreducibility of the gauge transformations (\ref{56}--\ref{59}). All these
results lead to the conclusion that the Koszul-Tate components $\delta _1$
and $\delta _2$ furnish a correct biresolution of $C^\infty (\Sigma ^{\prime
})$, where $\Sigma ^{\prime }$ is defined by the equations $\frac{\delta S_0%
}{\delta \Phi ^{\alpha _0}}=0$, such that it is permissible to approach the
antifield BRST-anti-BRST quantization of the irreducible gauge theory along
the general lines exposed in Section 2.

At this stage we construct the longitudinal exterior derivatives along the
gauge orbits, $D_1$ and $D_2$. In view of this we need to know the
coefficients appearing at the commutators among the irreducible gauge
transformations (\ref{56}--\ref{59}). As underlined in the previous
subsection, it is reasonable to assume the completeness of the irreducible
gauge generators. This assumption leads to the next general relations
expressing the on-shell closedness of the irreducible generators 
\begin{equation}
\label{4.39}Z_{\;\;\alpha _1}^{\alpha _0}\frac{\delta Z_{\;\;\beta
_1}^{\beta _0}}{\delta \Phi ^{\alpha _0}}-Z_{\;\;\beta _1}^{\alpha _0}\frac{%
\delta Z_{\;\;\alpha _1}^{\beta _0}}{\delta \Phi ^{\alpha _0}}\approx
C_{\;\;\alpha _1\beta _1}^{\gamma _1}Z_{\;\;\gamma _1}^{\beta _0}, 
\end{equation}
\begin{equation}
\label{4.40}Z_{\;\;\alpha _1}^{\alpha _0}\frac{\delta A_{\beta
_1}^{\;\;\alpha _2}}{\delta \Phi ^{\alpha _0}}-Z_{\;\;\beta _1}^{\alpha _0}%
\frac{\delta A_{\alpha _1}^{\;\;\alpha _2}}{\delta \Phi ^{\alpha _0}}\approx
C_{\;\;\alpha _1\beta _1}^{\gamma _1}A_{\gamma _1}^{\;\;\alpha
_2}+C_{\;\;\alpha _1\beta _1}^{\gamma _3}Z_{\;\;\gamma _3}^{\alpha _2}, 
\end{equation}
\begin{equation}
\label{4.41}Z_{\;\;\alpha _1}^{\alpha _0}\frac{\delta Z_{\;\;\beta
_{2k+1}}^{\alpha _{2k}}}{\delta \Phi ^{\alpha _0}}\approx C_{\;\;\alpha
_1\beta _{2k+1}}^{\gamma _{2k+1}}Z_{\;\;\gamma _{2k+1}}^{\alpha
_{2k}}+C_{\;\;\alpha _1\beta _{2k+1}}^{\gamma _{2k-1}}A_{\gamma
_{2k-1}}^{\;\;\alpha _{2k}},\;k=1,\cdots ,b, 
\end{equation}
\begin{equation}
\label{4.42}Z_{\;\;\alpha _1}^{\alpha _0}\frac{\delta A_{\beta
_{2k-1}}^{\;\;\alpha _{2k}}}{\delta \Phi ^{\alpha _0}}\approx C_{\;\;\alpha
_1\beta _{2k-1}}^{\gamma _{2k-1}}A_{\gamma _{2k-1}}^{\;\;\alpha
_{2k}}+C_{\;\;\alpha _1\beta _{2k-1}}^{\gamma _{2k+1}}Z_{\;\;\gamma
_{2k+1}}^{\alpha _{2k}},\;k=2,\cdots ,a, 
\end{equation}
where the coefficients involved with the right-hand sides of the prior
formulas may depend in principle on the fields $\Phi ^{\alpha _0}$.

The definitions of $D_1$ and $D_2$ acting on the generators (\ref{4.2}) are
constructed with the help of (\ref{2.12}--\ref{2.15}) and (\ref{4.39}--\ref
{4.42}). In the sequel we omit the superscript for notational simplicity.
Related to the ghost number zero fields, these definitions take the concrete
form 
\begin{equation}
\label{4.43}D_1\Phi ^{\alpha _0}=Z_{\;\;\alpha _1}^{\alpha _0}\eta
_1^{\alpha _1},\;D_2\Phi ^{\alpha _0}=Z_{\;\;\alpha _1}^{\alpha _0}\eta
_2^{\alpha _1}, 
\end{equation}
\begin{equation}
\label{4.44}D_1\Phi ^{\alpha _{2k}}=Z_{\;\;\alpha _{2k+1}}^{\alpha
_{2k}}\eta _1^{\alpha _{2k+1}}+A_{\alpha _{2k-1}}^{\;\;\alpha _{2k}}\eta
_1^{\alpha _{2k-1}},\;k=1,\cdots ,a, 
\end{equation}
\begin{equation}
\label{4.45}D_2\Phi ^{\alpha _{2k}}=Z_{\;\;\alpha _{2k+1}}^{\alpha
_{2k}}\eta _2^{\alpha _{2k+1}}+A_{\alpha _{2k-1}}^{\;\;\alpha _{2k}}\eta
_2^{\alpha _{2k-1}},\;k=1,\cdots ,a. 
\end{equation}
The actions of $D_1$ and $D_2$ on the ghosts are given by 
\begin{equation}
\label{4.46}D_1\eta _1^{\alpha _1}=\frac 12C_{\;\;\beta _1\gamma _1}^{\alpha
_1}\eta _1^{\beta _1}\eta _1^{\gamma _1}+C_{\;\;\beta _1\gamma _3}^{\alpha
_1}\eta _1^{\beta _1}\eta _1^{\gamma _3}, 
\end{equation}
\begin{eqnarray}\label{4.47}
& &D_1\eta _1^{\alpha _{2k+1}}=C_{\;\;\beta _1\gamma _{2k+1}}^{\alpha
_{2k+1}}\eta _1^{\beta _1}\eta _1^{\gamma _{2k+1}}+\nonumber \\
& &C_{\;\;\beta _1\gamma _{2k+3}}^{\alpha _{2k+1}}\eta _1^{\beta
_1}\eta _1^{\gamma _{2k+3}}+\bar a_{2k+1}C_{\;\;\beta _1\gamma
_{2k-1}}^{\alpha _{2k+1}}\eta _1^{\beta _1}\eta _1^{\gamma
_{2k-1}},\;k=1,\cdots ,b, 
\end{eqnarray}
\begin{equation}
\label{4.48}D_1\eta _2^{\alpha _1}=-\pi ^{\alpha _1}+\frac 12C_{\;\;\beta
_1\gamma _1}^{\alpha _1}\eta _1^{\beta _1}\eta _2^{\gamma _1}+\frac
12C_{\;\;\beta _1\gamma _3}^{\alpha _1}\left( \eta _2^{\beta _1}\eta
_1^{\gamma _3}+\eta _1^{\beta _1}\eta _2^{\gamma _3}\right) , 
\end{equation}
\begin{eqnarray}\label{4.49}
& &D_1\eta _2^{\alpha _{2k+1}}=-\pi ^{\alpha _{2k+1}}+\frac 12C_{\;\;\beta
_1\gamma _{2k+1}}^{\alpha _{2k+1}}\left( \eta _2^{\beta _1}\eta _1^{\gamma
_{2k+1}}+\eta _1^{\beta _1}\eta _2^{\gamma _{2k+1}}\right) +\nonumber \\ 
& &\frac 12C_{\;\;\beta _1\gamma _{2k+3}}^{\alpha _{2k+1}}%
\left( \eta _2^{\beta
_1}\eta _1^{\gamma _{2k+3}}+\eta _1^{\beta _1}\eta _2^{\gamma
_{2k+3}}\right) +\nonumber \\ 
& &\frac 12\bar a_{2k+1}C_{\;\;\beta _1\gamma _{2k-1}}^{\alpha
_{2k+1}}\left( \eta _2^{\beta _1}\eta _1^{\gamma _{2k-1}}+\eta _1^{\beta
_1}\eta _2^{\gamma _{2k-1}}\right) ,\;k=1,\cdots ,b, 
\end{eqnarray}
\begin{equation}
\label{4.50}D_2\eta _1^{\alpha _1}=\pi ^{\alpha _1}+\frac 12C_{\;\;\beta
_1\gamma _1}^{\alpha _1}\eta _1^{\beta _1}\eta _2^{\gamma _1}+\frac
12C_{\;\;\beta _1\gamma _3}^{\alpha _1}\left( \eta _1^{\beta _1}\eta
_2^{\gamma _3}+\eta _2^{\beta _1}\eta _1^{\gamma _3}\right) , 
\end{equation}
\begin{eqnarray}\label{4.51}
& &D_2\eta _1^{\alpha _{2k+1}}=\pi ^{\alpha _{2k+1}}+\frac 12C_{\;\;\beta
_1\gamma _{2k+1}}^{\alpha _{2k+1}}\left( \eta _1^{\beta _1}\eta _2^{\gamma
_{2k+1}}+\eta _2^{\beta _1}\eta _1^{\gamma _{2k+1}}\right) +\nonumber \\ 
& &\frac 12C_{\;\;\beta _1\gamma _{2k+3}}^{\alpha _{2k+1}}%
\left( \eta _1^{\beta
_1}\eta _2^{\gamma _{2k+3}}+\eta _2^{\beta _1}\eta _1^{\gamma
_{2k+3}}\right) +\nonumber \\ 
& &\frac 12\bar a_{2k+1}C_{\;\;\beta _1\gamma _{2k-1}}^{\alpha
_{2k+1}}\left( \eta _1^{\beta _1}\eta _2^{\gamma _{2k-1}}+\eta _2^{\beta
_1}\eta _1^{\gamma _{2k-1}}\right) ,\;k=1,\cdots ,b, 
\end{eqnarray}
\begin{equation}
\label{4.52}D_2\eta _2^{\alpha _1}=\frac 12C_{\;\;\beta _1\gamma _1}^{\alpha
_1}\eta _2^{\beta _1}\eta _2^{\gamma _1}+C_{\;\;\beta _1\gamma _3}^{\alpha
_1}\eta _2^{\beta _1}\eta _2^{\gamma _3}, 
\end{equation}
\begin{eqnarray}\label{4.53}
& &D_2\eta _2^{\alpha _{2k+1}}=C_{\;\;\beta _1\gamma _{2k+1}}^{\alpha
_{2k+1}}\eta _2^{\beta _1}\eta _2^{\gamma _{2k+1}}+\nonumber \\ 
& &C_{\;\;\beta _1\gamma _{2k+3}}^{\alpha _{2k+1}}\eta _2^{\beta
_1}\eta _2^{\gamma _{2k+3}}+\bar a_{2k+1}C_{\;\;\beta _1\gamma
_{2k-1}}^{\alpha _{2k+1}}\eta _2^{\beta _1}\eta _2^{\gamma
_{2k-1}},\;k=1,\cdots ,b. 
\end{eqnarray}
Finally, we have the following actions with respect to the ghosts of ghosts 
\begin{equation}
\label{4.54}D_1\pi ^{\alpha _1}=\frac 12C_{\;\;\beta _1\gamma _1}^{\alpha
_1}\pi ^{\beta _1}\eta _1^{\gamma _1}+\frac 12C_{\;\;\beta _1\gamma
_3}^{\alpha _1}\left( \pi ^{\beta _1}\eta _1^{\gamma _3}-\eta _1^{\beta
_1}\pi ^{\gamma _3}\right) , 
\end{equation}
\begin{eqnarray}\label{4.55}
& &D_1\pi ^{\alpha _{2k+1}}=\frac 12C_{\;\;\beta _1\gamma _{2k+1}}^{\alpha
_{2k+1}}\left( \pi ^{\beta _1}\eta _1^{\gamma _{2k+1}}-\eta _1^{\beta _1}\pi
^{\gamma _{2k+1}}\right) +\nonumber \\ 
& &\frac 12C_{\;\;\beta _1\gamma _{2k+3}}^{\alpha _{2k+1}}\left( \pi ^{\beta
_1}\eta _1^{\gamma _{2k+3}}-\eta _1^{\beta _1}\pi ^{\gamma _{2k+3}}\right) + 
\nonumber \\
& &\frac 12\bar a_{2k+1}C_{\;\;\beta _1\gamma _{2k-1}}^{\alpha
_{2k+1}}\left( \pi ^{\beta _1}\eta _1^{\gamma _{2k-1}}-\eta _1^{\beta _1}\pi
^{\gamma _{2k-1}}\right) ,\;k=1,\cdots ,b, 
\end{eqnarray}
\begin{equation}
\label{4.56}D_2\pi ^{\alpha _1}=\frac 12C_{\;\;\beta _1\gamma _1}^{\alpha
_1}\pi ^{\beta _1}\eta _2^{\gamma _1}+\frac 12C_{\;\;\beta _1\gamma
_3}^{\alpha _1}\left( \pi ^{\beta _1}\eta _2^{\gamma _3}-\eta _2^{\beta
_1}\pi ^{\gamma _3}\right) , 
\end{equation}
\begin{eqnarray}\label{4.57}
& &D_2\pi ^{\alpha _{2k+1}}=\frac 12C_{\;\;\beta _1\gamma _{2k+1}}^{\alpha
_{2k+1}}\left( \pi ^{\beta _1}\eta _2^{\gamma _{2k+1}}-\eta _2^{\beta _1}\pi
^{\gamma _{2k+1}}\right) +\nonumber \\ 
& &\frac 12C_{\;\;\beta _1\gamma _{2k+3}}^{\alpha _{2k+1}}\left( \pi ^{\beta
_1}\eta _2^{\gamma _{2k+3}}-\eta _2^{\beta _1}\pi ^{\gamma _{2k+3}}\right) +
\nonumber \\
& &\frac 12\bar a_{2k+1}C_{\;\;\beta _1\gamma _{2k-1}}^{\alpha
_{2k+1}}\left( \pi ^{\beta _1}\eta _2^{\gamma _{2k-1}}-\eta _2^{\beta _1}\pi
^{\gamma _{2k-1}}\right) ,\;k=1,\cdots ,b. 
\end{eqnarray}
In the above, we used the notation 
\begin{equation}
\label{4.58}\bar a_{2k+1}=\left\{ 
\begin{array}{c}
\frac 12,\; 
{\rm for}\;k=1, \\ 1,\;{\rm for}\;k\neq 1. 
\end{array}
\right. 
\end{equation}
Taking into account the above definitions of $\delta _1$, $\delta _2$, $D_1$
and $D_2$, the first two pieces of the solution to the master equation (\ref
{2.50}) for our irreducible gauge theory read as 
\begin{equation}
\label{4.59}\stackrel{(0)}{S}=S_0\left[ \Phi ^{\alpha _0}\right] , 
\end{equation}
\begin{eqnarray}\label{4.60}
& &\stackrel{(1)}{S}=
\Phi _{\alpha _0}^{*(1)}Z_{\;\;\alpha _1}^{\alpha _0}\eta
_1^{\alpha _1}+\sum\limits_{k=1}^a\left( \Phi _{\alpha
_{2k}}^{*(1)}Z_{\;\;\alpha _{2k+1}}^{\alpha _{2k}}\eta _1^{\alpha
_{2k+1}}+\Phi _{\alpha _{2k}}^{*(1)}A_{\alpha _{2k-1}}^{\;\;\alpha
_{2k}}\eta _1^{\alpha _{2k-1}}\right) +\nonumber \\ 
& &\Phi _{\alpha _0}^{*(2)}Z_{\;\;\alpha _1}^{\alpha _0}\eta
_2^{\alpha _1}+\sum\limits_{k=1}^a\left( \Phi _{\alpha
_{2k}}^{*(2)}Z_{\;\;\alpha _{2k+1}}^{\alpha _{2k}}\eta _2^{\alpha
_{2k+1}}+\Phi _{\alpha _{2k}}^{*(2)}A_{\alpha _{2k-1}}^{\;\;\alpha
_{2k}}\eta _2^{\alpha _{2k-1}}\right) . 
\end{eqnarray}
The third piece from $S$ contains, apart from the usual terms exposed in
Section 2, some supplementary terms that take into account the more complete
definitions (\ref{4.46}--\ref{4.57}), and is expressed by%
\begin{eqnarray}\label{4.61}
& &\stackrel{(2)}{S}=\sum\limits_{k=0}^b\left( \eta _{\alpha
_{2k+1}}^{*(21)}-\eta _{\alpha _{2k+1}}^{*(12)}+\bar \Phi _{\alpha
_{2k}}Z_{\;\;\alpha _{2k+1}}^{\alpha _{2k}}+\bar \Phi _{\alpha
_{2k+2}}A_{\alpha _{2k+1}}^{\;\;\alpha _{2k+2}}\right) \pi ^{\alpha
_{2k+1}}+\nonumber \\ 
& &\eta _{\alpha _1}^{*(11)}\left( \frac 12C_{\;\;\beta _1\gamma _1}^{\alpha
_1}\eta _1^{\beta _1}\eta _1^{\gamma _1}+C_{\;\;\beta _1\gamma _3}^{\alpha
_1}\eta _1^{\beta _1}\eta _1^{\gamma _3}\right) +\nonumber \\ 
& &\sum\limits_{k=1}^b\eta _{\alpha _{2k+1}}^{*(11)}\left( C_{\;\;\beta
_1\gamma _{2k+1}}^{\alpha _{2k+1}}\eta _1^{\beta _1}\eta _1^{\gamma
_{2k+1}}+\right. \nonumber \\ 
& &\left. C_{\;\;\beta _1\gamma _{2k+3}}^{\alpha _{2k+1}}\eta _1^{\beta _1}\eta
_1^{\gamma _{2k+3}}+\bar a_{2k+1}C_{\;\;\beta _1\gamma _{2k-1}}^{\alpha
_{2k+1}}\eta _1^{\beta _1}\eta _1^{\gamma _{2k-1}}\right) +\nonumber \\ 
& &\left( \eta _{\alpha _1}^{*(21)}+\eta _{\alpha _1}^{*(12)}\right) \left(
\frac 12C_{\;\;\beta _1\gamma _1}^{\alpha _1}\eta _1^{\beta _1}\eta
_2^{\gamma _1}+\frac 12C_{\;\;\beta _1\gamma _3}^{\alpha _1}\left( \eta
_2^{\beta _1}\eta _1^{\gamma _{2k+1}}+\eta _1^{\beta _1}\eta _2^{\gamma
_{2k+1}}\right) \right) +\nonumber \\ 
& &\sum\limits_{k=1}^b\left( \eta _{\alpha _{2k+1}}^{*(21)}+\eta _{\alpha
_{2k+1}}^{*(12)}\right) \left( \frac 12C_{\;\;\beta _1\gamma
_{2k+3}}^{\alpha _{2k+1}}\left( \eta _2^{\beta _1}\eta _1^{\gamma
_{2k+3}}+\eta _1^{\beta _1}\eta _2^{\gamma _{2k+3}}\right) +\right. 
\nonumber \\
& &\frac 12C_{\;\;\beta _1\gamma _{2k+1}}^{\alpha _{2k+1}}%
\left( \eta _2^{\beta
_1}\eta _1^{\gamma _{2k+1}}+\eta _1^{\beta _1}\eta _2^{\gamma
_{2k+1}}\right) +\nonumber \\ 
& &\left. \frac 12\bar a_{2k+1}C_{\;\;\beta _1\gamma _{2k-1}}^{\alpha
_{2k+1}}\left( \eta _2^{\beta _1}\eta _1^{\gamma _{2k-1}}+\eta _1^{\beta
_1}\eta _2^{\gamma _{2k-1}}\right) \right) +\nonumber \\ 
& &\eta _{\alpha _1}^{*(22)}\left( \frac 12C_{\;\;\beta _1\gamma _1}^{\alpha
_1}\eta _2^{\beta _1}\eta _2^{\gamma _1}+C_{\;\;\beta _1\gamma _3}^{\alpha
_1}\eta _2^{\beta _1}\eta _2^{\gamma _3}\right) +\nonumber \\ 
& &\sum\limits_{k=1}^b\eta _{\alpha _{2k+1}}^{*(22)}\left( C_{\;\;\beta
_1\gamma _{2k+1}}^{\alpha _{2k+1}}\eta _2^{\beta _1}\eta _2^{\gamma
_{2k+1}}+\right. \nonumber \\ 
& &\left. C_{\;\;\beta _1\gamma _{2k+3}}^{\alpha _{2k+1}}%
\eta _2^{\beta _1}\eta
_2^{\gamma _{2k+3}}+\bar a_{2k+1}C_{\;\;\beta _1\gamma _{2k-1}}^{\alpha
_{2k+1}}\eta _2^{\beta _1}\eta _2^{\gamma _{2k-1}}\right) + 
\nonumber \\
& &\pi _{\alpha _1}^{*(1)}\left( \frac 12C_{\;\;\beta _1\gamma _1}^{\alpha
_1}\pi ^{\beta _1}\eta _1^{\gamma _1}+\frac 12C_{\;\;\beta _1\gamma
_3}^{\alpha _1}\left( \pi ^{\beta _1}\eta _1^{\gamma _3}-\eta _1^{\beta
_1}\pi ^{\gamma _3}\right) \right) +\nonumber \\ 
& &\sum\limits_{k=1}^b\pi _{\alpha _{2k+1}}^{*(1)}\left( \frac 12C_{\;\;\beta
_1\gamma _{2k+1}}^{\alpha _{2k+1}}\left( \pi ^{\beta _1}\eta _1^{\gamma
_{2k+1}}-\eta _1^{\beta _1}\pi ^{\gamma _{2k+1}}\right) +\right. \nonumber \\ 
& &\frac 12C_{\;\;\beta _1\gamma _{2k+3}}^{\alpha _{2k+1}}%
\left( \pi ^{\beta
_1}\eta _1^{\gamma _{2k+3}}-\eta _1^{\beta _1}\pi ^{\gamma _{2k+3}}\right) + 
\nonumber \\
& &\left. \frac 12\bar a_{2k+1}C_{\;\;\beta _1\gamma _{2k-1}}^{\alpha
_{2k+1}}\left( \pi ^{\beta _1}\eta _1^{\gamma _{2k-1}}-\eta _1^{\beta _1}\pi
^{\gamma _{2k-1}}\right) \right) +\nonumber \\ 
& &\pi _{\alpha _1}^{*(2)}\left( \frac 12C_{\;\;\beta _1\gamma _1}^{\alpha
_1}\pi ^{\beta _1}\eta _2^{\gamma _1}+\frac 12C_{\;\;\beta _1\gamma
_3}^{\alpha _1}\left( \pi ^{\beta _1}\eta _2^{\gamma _3}-\eta _2^{\beta
_1}\pi ^{\gamma _3}\right) \right) +\nonumber \\ 
& &\sum\limits_{k=1}^b\pi _{\alpha _{2k+1}}^{*(2)}\left( \frac 12C_{\;\;\beta
_1\gamma _{2k+1}}^{\alpha _{2k+1}}\left( \pi ^{\beta _1}\eta _2^{\gamma
_{2k+1}}-\eta _2^{\beta _1}\pi ^{\gamma _{2k+1}}\right) +\right. \nonumber \\ 
& &\frac 12C_{\;\;\beta _1\gamma _{2k+3}}^{\alpha _{2k+1}}\left( \pi ^{\beta
_1}\eta _2^{\gamma _{2k+3}}-\eta _2^{\beta _1}\pi ^{\gamma _{2k+3}}\right) +
\nonumber \\
& &\left. \frac 12\bar a_{2k+1}C_{\;\;\beta _1\gamma
_{2k-1}}^{\alpha _{2k+1}}\left( \pi ^{\beta _1}\eta _2^{\gamma _{2k-1}}-\eta
_2^{\beta _1}\pi ^{\gamma _{2k-1}}\right) \right) +\cdots . 
\end{eqnarray}
The remaining terms from $\stackrel{(2)}{S}$, as well as the higher-order
pieces of the solution to the master equation can be derived by means of
projecting the master equation in the antifield BRST-anti-BRST formalism on
increasing biresolution degrees.

The gauge-fixing process goes as explained in Section 2, and requires the
supplementary variables 
\begin{equation}
\label{4.62}\mu _{(1)}^A=\left( \stackrel{(0,1)}{\mu }_{(1)}^{(\Phi )\alpha
_{2k}},\stackrel{(1,1)}{\mu }_{(1)}^{(\eta _1)\alpha _{2k+1}},\stackrel{(0,2)%
}{\mu }_{(1)}^{(\eta _2)\alpha _{2k+1}},\stackrel{(1,2)}{\mu }_{(1)}^{(\pi
)\alpha _{2k+1}}\right) .
\end{equation}
With the help of these new fields, we pass to the solution of the master
equation (in the first antibracket) 
\begin{eqnarray}\label{4.63}
& &S_1=S+\int d^Dx\left( \sum\limits_{k=0}^a\Phi _{\alpha _{2k}}^{*(2)}\mu
_{(1)}^{(\Phi )\alpha _{2k}}+\right. \nonumber \\ 
& &\left. \sum\limits_{k=0}^b\left( \eta _{\alpha
_{2k+1}}^{*(21)}\mu _{(1)}^{(\eta _1)\alpha _{2k+1}}+\eta _{\alpha
_{2k+1}}^{*(22)}\mu _{(1)}^{(\eta _2)\alpha _{2k+1}}+\pi _{\alpha
_{2k+1}}^{*(2)}\mu _{(1)}^{(\pi )\alpha _{2k+1}}\right) \right) . 
\end{eqnarray}
The gauge-fixed action results as in Section 2 by choosing an appropriate
gauge-fixing boson. A possible gauge-fixing boson is expressed by 
\begin{equation}
\label{4.64}F=\frac 12\sum\limits_{k=0}^a\int d^Dx\left( \Phi ^{\alpha
_{2k}}K_{\alpha _{2k}\beta _{2k}}\Phi ^{\beta _{2k}}\right) ,
\end{equation}
where $K_{\alpha _{2k}\beta _{2k}}$ stand for some symmetric
field-independent invertible matrices playing the role of metric tensors
with respect to the field indices. Eliminating the bar variables and the
antifields conjugated with the `fields' in the first antibracket from (\ref
{4.63}) on behalf of (\ref{4.64}) (see (\ref{2.62})), and further all $\Phi
_A^{*(2)}$ and $\mu _{(1)}^A$ on their equations of motion, we consequently
arrive at the gauge-fixed action 
\begin{eqnarray}\label{4.65}
& &S_{1_F}=\int d^Dx\left( \sum\limits_{k=0}^b\left( \Phi ^{\beta
_{2k}}K_{\beta _{2k}\alpha _{2k}}Z_{\;\;\alpha _{2k+1}}^{\alpha _{2k}}+\Phi
^{\beta _{2k+2}}K_{\beta _{2k+2}\alpha _{2k+2}}A_{\alpha
_{2k+1}}^{\;\;\alpha _{2k+2}}\right) \pi ^{\alpha _{2k+1}}-\right. 
\nonumber \\
& &\left( Z_{\;\;\alpha _1}^{\alpha _0}\eta _2^{\alpha _1}\right) K_{\alpha
_0\beta _0}\left( Z_{\;\;\beta _1}^{\beta _0}\eta _1^{\beta _1}\right)
-\sum\limits_{k=1}^a\left( A_{\alpha _{2k-1}}^{\;\;\alpha _{2k}}\eta
_2^{\alpha _{2k-1}}+Z_{\;\;\alpha _{2k+1}}^{\alpha _{2k}}\eta _2^{\alpha
_{2k+1}}\right) K_{\alpha _{2k}\beta _{2k}}\times \nonumber \\ 
& &\left. \times \left( A_{\beta _{2k-1}}^{\;\;\beta _{2k}}\eta
_1^{\beta _{2k-1}}+Z_{\;\;\beta _{2k+1}}^{\beta _{2k}}\eta _1^{\beta
_{2k+1}}\right) \right) +S_0\left[ \Phi ^{\alpha _0}\right] +\cdots . 
\end{eqnarray}
If one eventually needs to enforce some Gaussian averages with respect to $%
\pi ^{\alpha _{2k+1}}$, then it is necessary to add to $F$ some terms
quadratic in the ghosts $\eta _1^{\alpha _{2k+1}}$ and $\eta _2^{\alpha
_{2k+1}}$. In this way, our irreducible treatment for reducible gauge
theories is completely elucidated.
	
\section{Examples}
\noindent
In this section we apply the general theory on two interesting models of
field theory, namely, the Freedman-Townsend model and an example involving
abelian three-form gauge fields.

\subsection{The Freedman-Townsend model}
\noindent
We start with the Lagrangian action
\begin{equation}
\label{ft1}S_0^L\left[ B_{\mu \nu }^a,A_\mu ^a\right] =\frac 12\int
d^4x\left( -B_a^{\mu \nu }F_{\mu \nu }^a+A_\mu ^aA_a^\mu \right) , 
\end{equation}
where $B_{\mu \nu }^a$ stands for an antisymmetric tensor field, and the
field strength, $F_{\mu \nu }^a$, is defined by 
\begin{equation}
\label{ft2}F_{\mu \nu }^a=\partial _\mu A_\nu ^a-\partial _\nu A_\mu
^a-f_{\;\;bc}^aA_\mu ^bA_\nu ^c. 
\end{equation}
Action (\ref{ft1}) is invariant under the first-stage reducible gauge
transformations 
\begin{equation}
\label{ft3}\delta _\epsilon B_{\mu \nu }^a=\varepsilon _{\mu \nu \lambda
\rho }\left( D^\lambda \right) _{\;\;b}^a\epsilon ^{\rho b},\;\delta
_\epsilon A_\mu ^a=0, 
\end{equation}
with 
\begin{equation}
\label{ft4}\left( D^\lambda \right) _{\;\;b}^a=\delta _{\;\;b}^a\partial
^\lambda +f_{\;\;bc}^aA^{\lambda c}. 
\end{equation}
The field equations deriving from (\ref{ft1}) read as 
\begin{equation}
\label{ft5}\frac{\delta S_0^L}{\delta B_{\mu \nu }^a}\equiv -\frac
12F_a^{\mu \nu }=0,\;\frac{\delta S_0^L}{\delta A_a^\mu }\equiv A_\mu
^a+\left( D^\lambda \right) _{\;\;b}^aB_{\lambda \mu }^b=0. 
\end{equation}
The non-vanishing gauge generators of (\ref{ft3}) 
\begin{equation}
\label{ft6}\left( Z_{\mu \nu \rho }\right) _{\;\;b}^a=\varepsilon _{\mu \nu
\lambda \rho }\left( D^\lambda \right) _{\;\;b}^a, 
\end{equation}
admit the first-order on-shell reducibility relations 
\begin{equation}
\label{ft7}\left( Z_{\mu \nu \rho }\right) _{\;\;b}^a\left( Z^\rho \right)
_{\;\;c}^b=-\frac 12\varepsilon _{\mu \nu \lambda \rho }f_{\;\;cd}^a\frac{%
\delta S_0^L}{\delta B_{\lambda \rho d}}, 
\end{equation}
where the first-stage reducibility functions are expressed by 
\begin{equation}
\label{ft8}\left( Z^\rho \right) _{\;\;c}^b=\left( D^\rho \right)
_{\;\;c}^b. 
\end{equation}

To every reducibility relation (\ref{ft7}) we attach a scalar field, $%
\varphi ^{\alpha _2}\equiv \varphi ^c$, subject to the gauge transformations 
\begin{equation}
\label{ft14}\delta _\epsilon \varphi ^{\alpha _2}=A_{\;\;\alpha _1}^{\alpha
_2}\epsilon ^{\alpha _1}, 
\end{equation}
where $A_{\;\;\alpha _1}^{\alpha _2}$ is such that $A_{\;\;\alpha _1}^{\beta
_2}Z_{\;\;\alpha _2}^{\alpha _1}$ is invertible. For example, we take 
\begin{equation}
\label{ft16}A_{\;\;\alpha _1}^{\beta _2}=-\delta _{\;\;b}^a\partial _\rho
^y\delta ^4\left( x-y\right) , 
\end{equation}
hence 
\begin{equation}
\label{ft18}\delta _\epsilon \varphi ^a=\partial ^\mu \epsilon _\mu ^a. 
\end{equation}

The field, ghost, and antifield spectra of the antifield BRST-anti-BRST
background are respectively given by 
\begin{equation}
\label{ft40}\left( \stackrel{(0,0)}{B}_{\mu \nu }^a,\stackrel{(0,0)}{A}_\mu
^a,\stackrel{(0,0)}{\varphi }^a,\stackrel{(1,0)}{\eta }_1^{\nu a},\stackrel{%
(0,1)}{\eta }_2^{\nu a},\stackrel{(1,1)}{\pi }^{\nu a}\right) , 
\end{equation}
\begin{equation}
\label{ft41}\left( \stackrel{(-1,0)}{B}_a^{*(1)\mu \nu },\stackrel{(-1,0)}{A}%
_a^{*(1)\mu },\stackrel{(-1,0)}{\varphi }_a^{*(1)},\stackrel{(-2,0)}{\eta }%
_{\nu a}^{*(11)},\stackrel{(-1,-1)}{\eta }_{\nu a}^{*(12)},\stackrel{(-2,-1)%
}{\pi }_{\nu a}^{*(1)}\right) , 
\end{equation}
\begin{equation}
\label{ft42}\left( \stackrel{(0,-1)}{B}_a^{*(2)\mu \nu },\stackrel{(0,-1)}{A}%
_a^{*(2)\mu },\stackrel{(0,-1)}{\varphi }_a^{*(2)},\stackrel{(-1,-1)}{\eta }%
_{\nu a}^{*(21)},\stackrel{(0,-2)}{\eta }_{\nu a}^{*(22)},\stackrel{(-1,-2)}{%
\pi }_{\nu a}^{*(2)}\right) , 
\end{equation}
\begin{equation}
\label{ft43}\left( \stackrel{(-1,-1)}{\bar B}_a^{\mu \nu },\stackrel{(-1,-1)%
}{\bar A}_a^\mu ,\stackrel{(-1,-1)}{\bar \varphi }_a,\stackrel{(-2,-1)}{\bar
\eta }_{1\nu a},\stackrel{(-1,-2)}{\bar \eta }_{2\nu a},\stackrel{(-2,-2)}{%
\bar \pi }_{\nu a}\right) . 
\end{equation}
In the following we discard the superscript for the sake of notation
simplicity. The solution of the master equation associated with the
irreducible formalism takes the form%
\begin{eqnarray}\label{ft46}
& &\bar S=S_0^L+\int d^4x\left( \varepsilon _{\mu \nu \lambda \rho }\left(
B_a^{*(1)\mu \nu }\left( D^\lambda \right) _{\;\;b}^a\eta _1^{\rho
b}+B_a^{*(2)\mu \nu }\left( D^\lambda \right) _{\;\;b}^a\eta _2^{\rho
b}\right) +\right. \nonumber \\  
& &\varphi _a^{*(1)}\partial ^\mu \eta _{1\mu }^a+\varphi _a^{*(2)}\partial
^\mu \eta _{2\mu }^a+\left( \eta _{\nu a}^{*(21)}-\eta _{\nu
a}^{*(12)}\right) \pi ^{\nu a}+\nonumber \\ 
& &\left. \varepsilon _{\mu \nu \lambda \rho }\bar B_a^{\mu \nu
}\left( D^\lambda \right) _{\;\;b}^a\pi ^{\rho b}+\bar \varphi _a\partial
^\mu \pi _\mu ^a\right) .
\end{eqnarray}
In connection with the gauge-fixing procedure, we introduce the fields 
\begin{equation}
\label{ft47}\left( \stackrel{(0,1)}{\mu }_{(1)\mu \nu }^{(B)a},\stackrel{%
(0,1)}{\mu }_{(1)\mu }^{(A)a},\stackrel{(0,1)}{\mu }_{(1)}^{(\varphi )a},%
\stackrel{(1,1)}{\mu }_{(1)\mu }^{(\eta _1)a},\stackrel{(0,2)}{\mu }_{(1)\mu
}^{(\eta _2)a},\stackrel{(1,2)}{\mu }_{(1)\mu }^{(\pi )a}\right) , 
\end{equation}
and work with the solution 
\begin{eqnarray}\label{ft48}
& &S_1=\bar S+\int d^4x\left( B_a^{*(2)\mu \nu }\mu _{(1)\mu \nu
}^{(B)a}+A_a^{*(2)\mu }\mu _{(1)\mu }^{(A)a}+\varphi _a^{*(2)}\mu
_{(1)}^{(\varphi )a}+\right. \nonumber \\ 
& &\left. \eta _{\mu a}^{*(21)}\mu _{(1)}^{(\eta _1)\mu a}+\eta
_{\mu a}^{*(22)}\mu _{(1)}^{(\eta _2)\mu a}+\pi _{\mu a}^{*(2)}\mu
_{(1)}^{(\pi )\mu a}\right) . 
\end{eqnarray}
Choosing the gauge-fixing boson of the type (\ref{4.64}) plus an appropriate
term that leads to a Gaussian average, namely, 
\begin{equation}
\label{ft49}F=\int d^4x\left( -\frac 14B_{\mu \nu }^aB_a^{\mu \nu }+\frac
12\varphi ^a\varphi _a+\eta _{2\nu a}\eta _1^{\nu a}\right) , 
\end{equation}
eliminating the antifields with the index $(1)$ and the bar variables in the
standard way (with the help of (\ref{ft49})) from $S_1$, and subsequently
eliminating the antifields bearing the index $(2)$ and the $\mu _{(1)}$'s on
their equations of motion, we finally reach the gauge-fixed action%
\begin{eqnarray}\label{ft50}
& &S_{1_F}=S_0^L+\int d^4x\left( -\frac 12\left( \left( D_{\left[ \lambda
\right. }\right) _{\;\;a}^d\eta _{2\left. \rho \right] d}\right) \left(
\left( D^{\left[ \lambda \right. }\right) _{\;\;b}^a\eta _1^{\left. \rho
\right] b}\right) -\right. \nonumber \\  
& &\left. \left( \partial ^\rho \eta _{2\rho a}\right) \left(
\partial _\mu \eta _1^{\mu a}\right) +\left( \frac 12\varepsilon _{\mu \nu
\lambda \rho }\left( D^\mu \right) _{\;\;a}^bB_b^{\nu \lambda }-\partial
_\rho \varphi _a-\pi _{\rho a}\right) \pi _a^\rho \right) .
\end{eqnarray}
Action (\ref{ft50}) possesses no gauge invariances. This ends our
irreducible antifield BRST-anti-BRST treatment of the Freedman-Townsend
model.

\subsection{A model with three-form gauge fields}
\noindent
Here, we start with the Lagrangian action
\begin{equation}
\label{fm1}S_0^L\left[ A_{\mu \nu \lambda }\right] =\int d^7x\left( \frac{%
2\alpha ^2\left( 3!\right) ^24!}{M^2}F_{\mu \nu \lambda \rho }F^{\mu \nu
\lambda \rho }+\alpha \varepsilon _{\mu \nu \lambda \rho \sigma \beta \gamma
}F^{\mu \nu \lambda \rho }A^{\sigma \beta \gamma }\right) , 
\end{equation}
where $A_{\mu \nu \lambda }$ denote abelian three-form gauge fields, $%
\varepsilon _{\mu \nu \lambda \rho \sigma \beta \gamma }$ stands for the
completely antisymmetric symbol in seven dimensions, $\alpha $ and $M$ are
some constants, and the field strength is defined by 
\begin{equation}
\label{fm2}F_{\mu \nu \lambda \rho }=\partial _\mu A_{\nu \lambda \rho
}-\partial _\nu A_{\mu \lambda \rho }+\partial _\lambda A_{\rho \mu \nu
}-\partial _\rho A_{\lambda \mu \nu }\equiv \partial _{\left[ \mu \right.
}A_{\left. \nu \lambda \rho \right] }. 
\end{equation}
Action (\ref{fm1}) is invariant under the gauge transformations 
\begin{equation}
\label{fm3}\delta _\epsilon A^{\mu \nu \lambda }=\partial ^{\left[ \mu
\right. }\epsilon ^{\left. \nu \lambda \right] }, 
\end{equation}
where the gauge generators are of the form 
\begin{equation}
\label{fm4}Z_{\;\;\beta \gamma }^{\mu \nu \lambda }=\frac 12\partial
^{\left[ \mu \right. }\delta _{\;\;\beta }^\nu \delta _{\;\;\gamma }^{\left.
\lambda \right] }. 
\end{equation}
The above gauge generators are second stage reducible, with the reducibility
relations 
\begin{equation}
\label{fm5}Z_{\;\;\beta \gamma }^{\mu \nu \lambda }Z_{\;\;\rho }^{\beta
\gamma }=0, 
\end{equation}
\begin{equation}
\label{fm6}Z_{\;\;\rho }^{\beta \gamma }Z^\rho =0, 
\end{equation}
where the first, respectively, second order reducibility functions are
expressed by 
\begin{equation}
\label{fm7}Z_{\;\;\rho }^{\beta \gamma }=\partial ^{\left[ \beta \right.
}\delta _{\;\;\rho }^{\left. \gamma \right] }, 
\end{equation}
\begin{equation}
\label{fm8}Z^\rho =\partial ^\rho . 
\end{equation}
The role of the indices $\alpha _0$, $\alpha _1$, and $\alpha _2$ is played
in our case by $\mu \nu \lambda $, $\beta \gamma $, respectively, $\rho $,
while $\alpha _3$ is one-valued and is omitted for notational simplicity. In
agreement with Section 3, we add the fields $A^{\alpha _2}\equiv A^\rho $,
the gauge parameters $\epsilon ^{\alpha _3}\equiv \epsilon $, and impose the
gauge transformations of the new fields like 
\begin{equation}
\label{fm9}\delta _\epsilon A^\rho =\partial _\lambda \epsilon ^{\lambda
\rho }+\partial ^\rho \epsilon . 
\end{equation}
It is clear that the gauge transformations (\ref{fm3}) and (\ref{fm9}) form
a complete and irreducible set. The field, ghost and antifield spectra are
organized as 
\begin{equation}
\label{fm10}\left( \stackrel{(0,0)}{A}^{\mu \nu \lambda },\stackrel{(0,0)}{A}%
^\mu ,\stackrel{(1,0)}{\eta }_1^{\mu \nu },\stackrel{(1,0)}{\eta }_1,%
\stackrel{(0,1)}{\eta }_2^{\mu \nu },\stackrel{(0,1)}{\eta }_2,\stackrel{%
(1,1)}{\pi }^{\mu \nu },\stackrel{(1,1)}{\pi }\right) , 
\end{equation}
\begin{eqnarray}\label{fm11}
& &\left( \stackrel{(-1,0)}{A}_{\mu \nu \lambda }^{*(1)},\stackrel{(-1,0)}{A}%
_\mu ^{*(1)},\stackrel{(-2,0)}{\eta }_{\mu \nu }^{*(11)},
\stackrel{(-2,0)}{\eta }^{*(11)},\right. \nonumber \\ 
& &\left. \stackrel{(-1,-1)}{\eta }_{\mu \nu }^{*(12)},
\stackrel{(-1,-1)}{\eta }^{*(12)},
\stackrel{(-2,-1)}{\pi }_{\mu \nu }^{*(1)},
\stackrel{(-2,-1)}{\pi }^{*(1)}\right) , 
\end{eqnarray}
\begin{eqnarray}\label{fm12}
& &\left( \stackrel{(0,-1)}{A}_{\mu \nu \lambda }^{*(2)},\stackrel{(0,-1)}{A}%
_\mu ^{*(2)},\stackrel{(-1,-1)}{\eta }_{\mu \nu }^{*(21)},
\stackrel{(-1,-1)}{\eta }^{*(21)},\right. \nonumber \\
& &\left. \stackrel{(0,-2)}{\eta }_{\mu \nu }^{*(22)},
\stackrel{(0,-2)}{\eta }^{*(22)},
\stackrel{(-1,-2)}{\pi }_{\mu \nu }^{*(2)},
\stackrel{(-1,-2)}{\pi }^{*(2)}\right) , 
\end{eqnarray}
\begin{equation}
\label{fm13}\left( \stackrel{(-1,-1)}{\bar A}_{\mu \nu \lambda },\stackrel{%
(-1,-1)}{\bar A}_\mu ,\stackrel{(-2,-1)}{\bar \eta }_{1\mu \nu },\stackrel{%
(-2,-1)}{\bar \eta }_1,\stackrel{(-1,-2)}{\bar \eta }_{2\mu \nu },\stackrel{%
(-1,-2)}{\bar \eta }_2,\stackrel{(-2,-2)}{\bar \pi }_{\mu \nu },\stackrel{%
(-2,-2)}{\bar \pi }\right) . 
\end{equation}
The solution of the master equation (\ref{2.50}) reads as 
\begin{eqnarray}\label{fm14}
& &S=S_0^L+\int d^7x\left( A_{\mu \nu \lambda }^{*(1)}\partial ^{\left[ \mu
\right. }\eta _1^{\left. \nu \lambda \right] }+A_{\mu \nu \lambda
}^{*(2)}\partial ^{\left[ \mu \right. }\eta _2^{\left. \nu \lambda \right]
}+A_\rho ^{*(1)}\left( \partial _\lambda \eta _1^{\lambda \rho }+\partial
^\rho \eta _1\right) +\right. \nonumber \\
& &A_\rho ^{*(2)}\left( \partial _\lambda \eta _2^{\lambda \rho }+\partial
^\rho \eta _2\right) +\left( \eta _{\mu \nu }^{*(21)}-\eta _{\mu \nu
}^{*(12)}\right) \pi ^{\mu \nu }+\left( \eta ^{*(21)}-\eta ^{*(12)}\right)
\pi +\nonumber \\ 
& &\left. \bar A_{\mu \nu \lambda }\partial ^{\left[ \mu \right.
}\pi ^{\left. \nu \lambda \right] }+\bar A_\rho \left( \partial _\lambda \pi
^{\lambda \rho }+\partial ^\rho \pi \right) \right) . 
\end{eqnarray}
In order to fix the gauge, it is necessary to introduce the `fields' 
\begin{eqnarray}\label{fm15}
& &\left( \stackrel{(0,1)}{\mu }_{(1)}^{(A)\mu \nu \lambda },%
\stackrel{(0,1)}{\mu }_{(1)}^{(A)\mu },
\stackrel{(1,1)}{\mu }_{(1)}^{(\eta _1)\mu \nu },%
\stackrel{(1,1)}{\mu }_{(1)}^{(\eta _1)},\right. \nonumber \\ 
& &\left. \stackrel{(0,2)}{\mu }_{(1)}^{(\eta _2)\mu \nu },%
\stackrel{(0,2)}{\mu }_{(1)}^{(\eta _2)},\stackrel{(1,2)}{\mu }_{(1)}^{(\pi
)\mu \nu },\stackrel{(1,2)}{\mu }_{(1)}^{(\pi )}\right) , 
\end{eqnarray}
and to work with the solution%
\begin{eqnarray}\label{fm16}
& &S_1=S+\int d^7x\left( A_{\mu \nu \lambda }^{*(2)}\mu _{(1)}^{(A)\mu \nu
\lambda }+A_\mu ^{*(2)}\mu _{(1)}^{(A)\mu }+\eta _{\mu \nu }^{*(21)}\mu
_{(1)}^{(\eta _1)\mu \nu }+\eta ^{*(21)}\mu _{(1)}^{(\eta _1)}+\right. 
\nonumber \\
& &\left. \eta _{\mu \nu }^{*(22)}\mu _{(1)}^{(\eta _2)\mu \nu
}+\eta ^{*(22)}\mu _{(1)}^{(\eta _2)}+\pi _{\mu \nu }^{*(2)}\mu _{(1)}^{(\pi
)\mu \nu }+\pi ^{*(2)}\mu _{(1)}^{(\pi )}\right) . 
\end{eqnarray}
We choose the gauge-fixing boson also of the form (\ref{4.64}), i.e., 
\begin{equation}
\label{fm17}F=-\int d^7x\left( \frac 16A_{\mu \nu \lambda }A^{\mu \nu
\lambda }+\frac 12A_\mu A^\mu \right) , 
\end{equation}
and consequently derive the gauge-fixed action (after elimination of some
auxiliary `fields' on their equations of motion)%
\begin{eqnarray}\label{fm18}
& &S_{1_F}=S_0^L+\int d^7x\left( -\eta _{2\mu \nu }\Box \eta _1^{\mu \nu }-
\eta _2\Box \eta _1+\right. \nonumber \\ 
& &\left. \pi _{\mu \nu }\left( \partial _\lambda A^{\lambda \mu
\nu }+\frac 12\partial ^{\left[ \mu \right. }A^{\left. \nu \right] }\right)
+\pi \partial _\mu A^\mu \right) . 
\end{eqnarray}
It is clear that the gauge-fixed action displays a propagating character.
The same line can be applied if one adds to the action (\ref{fm1}) any
interaction terms that are invariant also under the gauge transformations (%
\ref{fm3}).
	
\section{Conclusion}
\noindent
To conclude with, in this paper we develop a method that allows the
application of the irreducible antifield BRST-anti-BRST quantization to a
large class of reducible gauge theories. The crucial point of our procedure
is expressed by the replacement of the starting reducible system by an
equivalent irreducible one, such that we can substitute the antifield
BRST-anti-BRST quantization of the reducible theory with that of
the corresponding irreducible system. The quantization of the irreducible
system follows the standard rules of the irreducible antifield
BRST-anti-BRST method, the acyclicity of the Koszul-Tate bicomplex being
ensured. In due course we emphasize a possible class of gauge-fixing bosons
which is relevant in the context of our procedure. Finally, we show how our
mechanism can be applied to practical solutions on two models of interest.

\end{document}